\documentclass[12pt]{article}

\usepackage{xcolor}
\usepackage{amsmath}
\usepackage{graphicx}
\usepackage{verbatim}
\usepackage{color}
\usepackage{subfigure}
\usepackage{hyperref}
\usepackage{verbatim} 
\usepackage{multirow}
\usepackage[normalem]{ulem}
\usepackage{authblk}

\title{Aqueous Amino Acids and Proteins Near the Surface of Gold in Hydrophilic and Hydrophobic Force Fields}
\author[1]{Grzegorz Nawrocki}
\author[2]{Marek Cieplak}
\affil[1,2]{Institute of Physics, Polish Academy of Sciences, Al. Lotnik\'ow 32/46, 02-668 Warsaw, Poland}

\begin{document}

\maketitle


\begin{abstract}
We calculate potentials of the mean force for
twenty amino acids in the vicinity of the (111) surface of gold, 
for several dipeptides, and for some analogs of the side chains, 
using molecular dynamics simulations
and the umbrella sampling method. We compare results
obtained within three different force fields: one hydrophobic
(for a contaminated surface) and two hydrophilic. All of these
fields lead to good binding with very different specificities and
different patterns in the density and polarization of water.
The covalent bond with the sulfur atom on cysteine
is modeled by the Morse potential. 
We demonstrate that binding energies of dipeptides are different
than the combined binding energies of their amino-acidic components.
For the hydrophobic gold, adsorption events of a small protein
are driven by attraction to the strongest binding amino acids.
This is not so in the hydrophilic cases -- a result of smaller
specificities combined with the difficulty for proteins, but not
for single amino acids, to penetrate the first layer of water.
The properties of water near the surface sensitively depend
on the force field.
\end{abstract}


\section{Introduction}

Interactions between proteins and solid surfaces are at the heart of
many potential applications in bionanotechnology \cite{hlady,Gray}
and medicine \cite{ballou_2004,rabias_2010}. One example is a 
response of a body to implants. The interactions are also
crucial for understanding of many phenomena that take place in 
nature such as bone formation, adhesion of mussels to rocks, 
and anti-freeze protection of animals \cite{cohavi_2009}.
The applications often involve nanoparticles \cite{Dawson}, either 
functionalized by ligands or not. The nanoparticles may or may not be toxic, 
depending on the interactions with the relevant proteins. A broader
issue is specificity: which amino acids (AA), peptides, or proteins
bind to a given solid in aqueous solutions well and which do not.

Procedures for synthesis of colloidal gold have been known since 
the times of Faraday \cite{Faraday}. This may explain, together with 
gold's biocompatibility, why nanoparticles of gold have been
particularly well studied, also in the context of interactions
with proteins \cite{Brown,Daniel,boisselier_2009}.
One of the goals of these studies is to harness optical
and plasmonic properties of such nanoparticles. Furthermore,
flat gold surfaces are used 
in single molecule spectroscopy studies of biomolecules
as specimen discs and gold-coated AFM tips because of
cysteine's ability to form covalent bonds with gold
which facilitates stretching \cite{Gaub,Nagy,Progress}.

Naturally, the subject of protein-gold interactions in water has also been
studied theoretically quite extensively through classical all atom simulations
\cite{heinz_2008,hoefling_2010_b,hoefling_2010_a,hoefling_2011,verde_2009,verde_2011,bizzarri_2003,bizzarri_2003b,iori_2009}.
There are also density functional theory studies for small organic
molecules without water or with a few frozen molecules of water
\cite{calzolari_2010,iori_2008,felice_2004,Nazmutdinov_2006}. 
The first principle studies include quantum effects and give 
insights into the physics of the problem, but their practicality
is limited in situations with water solutions which involves 
many possible configurations of the system. On the other
hand, the classical simulational studies use distinct force fields
and are carried out for different proteins (or selected AAs) so it is
hard to make comparisons between them and even harder to reach 
understanding of theirs physics.
The force fields used seem to have similar parameters. However,
the crucial differences are contained in the
Lennard-Jones parameters for interactions of gold with water.
Our previous studies on ZnO \cite{nawrocki_2013} and ZnS \cite{nawrocki_2014}
indicate that the nature of the density profiling of water near the solid 
is an important determinant of how single AAs couple to the solid.
For instance, formation of an articulated first layer tends to screen
proteins from the electric field of the solid and thus weaken the coupling
and may also lead to a steric barrier.
Thus proper accounting for the density profiling of water is of the 
essence in the theory of interactions of biomolecules with a solid.

\begin{figure}[ht]
\begin{center}
\includegraphics[scale=0.3]{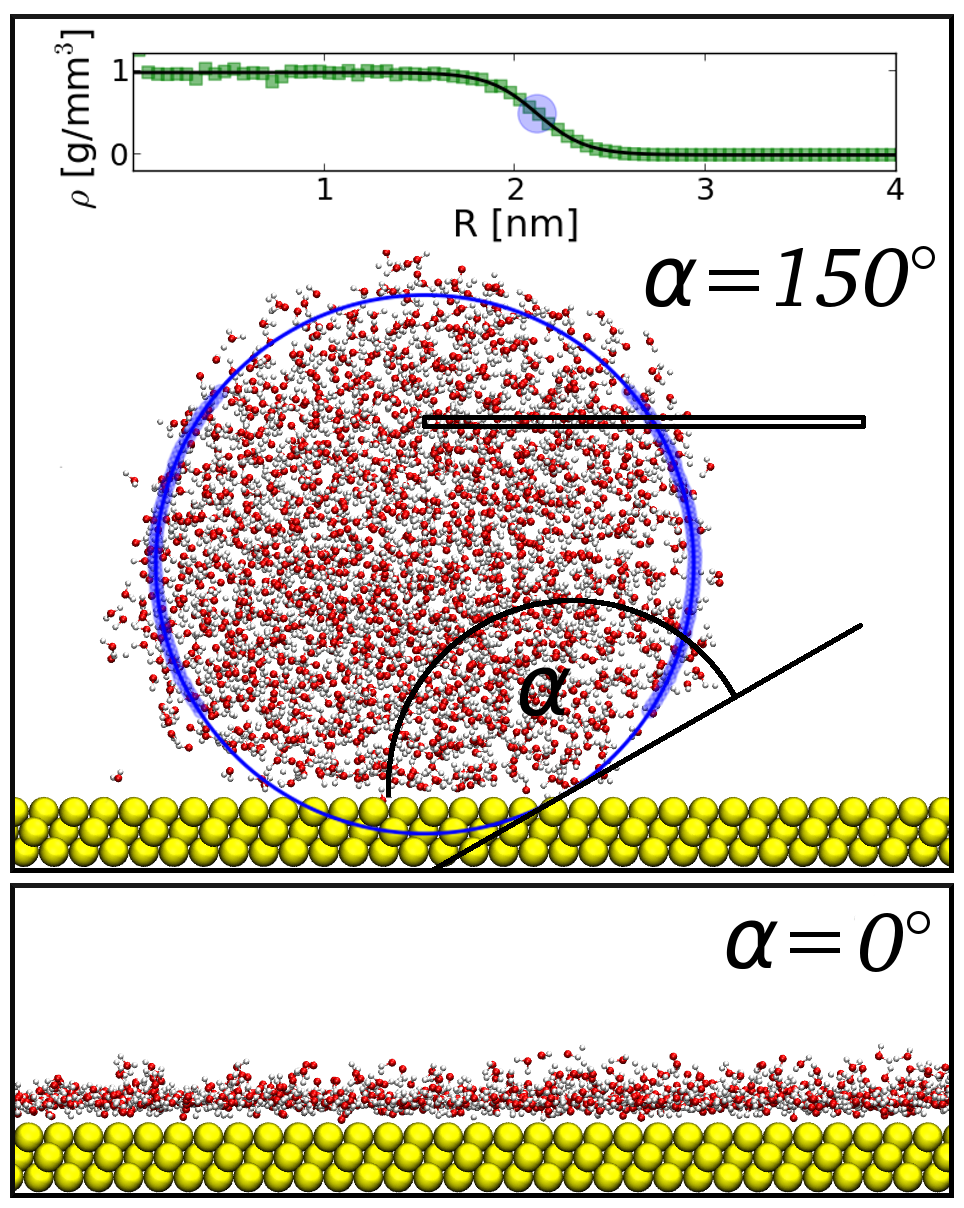}
\end{center}
  \caption{The final state of a droplet that is placed near the gold surface.
       The top panel is for FFB and the bottom for FFV.
       The initial state of the droplet is constructed by following
       the procedure of Verde {\it et al.} \cite{verde_2009}.
       The droplet is spherical initially and its diameter is set to 5 nm.
       Its lowest point is moved to 0.715 nm below the surface and all water
       molecules which are below the surface are removed, leaving about
       2000 molecules for further evolution. 
       The contact angle is determined by fitting the positions
       of atoms within the surface of the droplet into a circle
       obtained by axial averaging of the three-dimensional cluster.
       The regions at the top and bottom  of the circle (the thinner lines)
       are discarded from the fitting procedure.
       The plot at the top shows the average density of water 
       along a horizontal slab cut out from the droplet. 
       The inflection points (the blue circle) of the fitted sigmoidal 
       functions for many slabs determine 
       the surface of the droplet.}

\label{droplet}
\end{figure}

Here, we focus on three force fields: one hydrophobic of Bizzarri $et$ $al.$
\cite{bizzarri_2003,bizzarri_2003b}, henceforth denoted as FFB, and two
hydrophilic. The first of the latter has been proposed by
Verde $et$ $al.$ \cite{verde_2009,verde_2011} and will be denoted here as FFV.
Both FFB and FFV  are CHARMM-based \cite{mackerell_1998}  and all interactions
with gold are described by the Lennard-Jones potentials. The difference
between the two is captured by Figure \ref{droplet} which shows 
what happens to a bulk-equilibrated spherical droplet 
of about 2000 water molecules when placed near  the
Au (111) and then evolved using the GROMACS 4.5.5 package \cite{spoel_2005}.
For FFB, the sphere deforms into a droplet that makes the contact angle of
about 150$^{\circ}$ (the procedures will be described later).
On the other hand for FFV, there is perfect wetting: the molecular sphere
spreads evenly across the gold surface. The Lennard-Jones parameters in FFV
have been obtained by successive adjustments to make the initially finite
contact angle of the droplet just approach zero. The second hydrophilic
force field has been proposed by Iori {\it et al.} \cite{iori_2009}
and named GolP. Here, it will be referred to as FFI for uniformity
of notation for the three force fields considered.
The FFI field is hydrophilic, not because of the sufficiently strong 
van der Waals-based attraction but because it includes the
dynamical effects of charge polarization in gold.

The profound differences in the behavior of water between the
force fields must translate into
distinct energies of interactions of biomolecules with gold. Here, we 
shall bring this out by determining the potentials of the mean force (PMF) for
the twenty single AAs for FFB, FFV, and partially for FFI, and by demonstrating
their fundamental disparity. In particular, the binding energies of
AAs are, on average, about six times stronger in FFV compared
to FFB and the identities of the best non-cysteinic binders are distinct: 
ARG and MET for FFV with TRP and LEU for FFB. CYS is special as it
may form a covalent bond with Au. We model this bond
by using the Morse potential which allows for
attachment and detachment. This approach is qualitatively consistent 
with the experimental Raman spectra studies \cite{Finot,Finot1} which yield
the characteristic detachment time for cyteine to be of order 1 ms.
We then study the behavior of a small protein, the tryptophan cage,
near Au (111) and ask to what extent the knowledge of the single
AA potentials helps in interpreting binding events of proteins.
We also discuss interactions of gold with dipeptides and analogs
of AAs side chains.

The pertinent question to ask is which force field is closer to reality.
We think that actually both hydrophobic and hydrophilic fields may be useful.  
There are reports claiming that
gold is hydrophobic \cite{Apicella} and there are also reports
\cite{zisman_1965,Abdelsalam} that it is hydrophilic. The laboratory practice
shows \cite{Smith}  (see also ref. \cite{schrader_1984})
that clean gold is hydrophilic but it becomes
hydrophobic over time as the surface acquires carbon or oxygen contaminants.
Thus, fundamentally, the hydrophilic fields are more appropriate to use.
However, FFB can be considered as a useful effective parametrization
when the gold gets dirty. The rational design of sequences that bind well
to gold should then include information about the state of the surface,
but cysteine is found to bind strongly under whatever conditions.

It is a different matter to decide which of the two hydrophilic force
fields is more correct. 
In principle, it seems wiser to include the effects 
of the polarization of gold. However, the FFI model adopts a particular
way of doing it -- by invoking rotating dipoles that are placed
at non-obvious locations. It is not clear whether this empirical 
description is optimal.
There are alternative ways to take the delocalized electronic
charges in gold. One of them, for instance, 
employs the surface-integrated Lennard-Jones potentials in which 
the lateral corrugation part is reduced \cite{friction,friction1}
by an order of magnitude because of the delocalization effects.
This approach yields good agreement with experimental results on atomic
friction \cite{Krim}. Other approaches may involve adjustments in the
effective dielectric constant. It should also be noted that the
two hydrophilic force fields employ different models of the
molecules of water: in FFV, the atoms of H interact with Au
through Lennard-Jones interactions and in FFI they do not.

In any event, we find that the binding strengths  of single AAs
to gold are larger in FFV than in FFI by 
a factor of $1.47 \pm 0.41$ and that
FFV turns GLU and SER from relatively poor binders
under FFI into excellent binders while converting a good binder ARG
into the very top binder (other than CYS). Studying relationships 
between the binding strengths may offer an experimental way to
decide which force field to favor.
Another approach is to study the water density profiles
as a function of the vertical distance away from the surface.
We show that there are some shifts in the profiles 
between FFV and FFI. More clear-cut differences appear in the
distribution of water polarization just near the surface:
FFV favors flat orientations of the molecules of water in this
region whereas FFI -- staggered up and down. 

After discussing properties of water near gold and the PMFs for single
capped amino acids in the three force fields, we consider the
analogs of the AAs side groups and dipeptides. Finally, we analyze
adsorption events of a small protein -- the tryptophan cage --
and compare the results obtained with different force fields.
We find that the mechanisms of adsorption are distinct.

In addition to the force fields discussed in this paper, there
have been several other approaches proposed in the literature.
For instance, Braun $et$ $al.$ \cite{Braun} propose a $z$-dependent
10-4 potential with an amplitude determined through an order-of-magnitude
argument and combined with an electrostatically motivated corrugational term.
Heinz $et$ $al.$ \cite{heinz_2008,heinz_2009} consider an FFV-like force field
that does not include interactions with the hydrogen atoms in water molecules. 
Our selection of the force fields has been motivated by the desire to elucidate situations 
with the clearest and most distinct physical features. 
The hydrophilic properties of FFV are determined by the van der Waals interactions but in FFI by the polarization. 
FFV considers the polarization to be of no relevance 
whereas FFI assumes that Lennard-Jones interactions with the hydrogen atoms of water molecules to be not important. 
The selection of the hydrophobic FFB stems from the fact that FFB includes parameters 
describing the covalent Au-CYS interaction 
(it does not include the van der Waals interactions with hydrogen atoms in water molecules).

\section{Methods}

\subsection{Interactions and Force Fields}

Our all-atom MD simulations with explicit solvent are performed 
using the GROMACS 4.5.5 package \cite{spoel_2005}. 
The VMD package is used for visualization of structures \cite{VMD}.
The simulations involve both bonded and non-bonded interactions. 
The latter include Coulombic attraction and repulsion between charged particles, 
van der Waals forces, and overlapping of electronic orbitals. 
The latter two are described by the Lennard-Jones potentials. The bonded
interactions include penalties for stretching of covalent bonds
of length $b_{ij}^{0}$ (the term is $\frac{1}{2}k^b_{ij}(b_{ij}-b_{ij}^{0})^2$),
twisting of bond angles away from their preferred values
$\theta^0_{ijk}$ (the potential is $\frac{1}{2}k^{\theta}_{ijk}(\theta_{ijk}-\theta^0_{ijk})^2$,
bending of dihedral angles 
($k_{ijkl}^{\varphi}(1+cos(n\varphi_{ijkl}-\varphi_{ijkl}^{0}$)) in CHARMM 
and $\sum_{n=0}^5C_n(cos(\varphi_{ijkl}-180^o))^n$ in OPLS/AA),
and for not keeping aromatic rings planar 
($\frac{1}{2}k^\xi_{ijkl}(\xi_{ijkl}-\xi_{ijkl}^{0})^2$,
where $\xi_{ijkl}$ denotes the angle between planes set by atoms
$ijk$ and $jkl$,
and potentials like for the dihedral terms in OPLS/AA). 

The parametrization employed in FFB and FFV is basically CHARMM-based \cite{mackerell_1998}.
The distinction between the hydrophobic FFB and the hydrophilic FFV sits in 
the parameters associated with the non-bonded interactions with the surface
as described by the Lennard-Jones potential. The length parameter,
$\sigma_{AuAu}$, is 0.19 nm and 0.29 nm in FFB and FFV respectively.
This means that the atoms of Au in FFB act as smaller centers than in FFV.
The energy parameter associated with gold, 
$\epsilon_{AuAu}$, is 0.33 kJ/mol for FFB and 4.39 kJ/mol for FFV.
The FFB model \cite{bizzarri_2003,bizzarri_2003b} 
takes $\sigma_{AuAu}$ and $\epsilon_{AuAu}$
derived from the paper by Qian $et$ $al.$ \cite{qian_1997}.

The FFV model \cite{verde_2009,verde_2011} draws on 
the Universal Force Field \cite{rappe_1992} 
and then adjusts the parameters to reduce the contact angle to zero
when a droplet is placed on the surface. 

When using the FFB and FFV models we take water molecules
to be described by the TIP3P model \cite{jorgensen_1983}.
However, in the case of FFV we use its modified version, 
proposed by Verde $et$ $al.$ \cite{verde_2009}, in which not only the
oxygen but also the hydrogen atoms interact with gold and all other
components of the system through the Lennard-Jones potentials (in addition
to the Coulombic terms). 
An alternative approach \cite{heinz_2008} is to eliminate the Lennard-Jones
couplings with the hydrogen atoms and enhance $\epsilon_{AuAu}$ to about 22 kJ/mol.

In the FFI force field \cite{iori_2009}, the relevant parameters 
are taken from the OPLS/AA force field \cite{jorgensen_1996}
and water is described by the SPC model \cite{berendsen_1981}.
However, the Au surface is considered to be polarizable.
The polarization is taken into account by placing permanent electric
dipoles of length 0.07 nm and charges $\pm$ 0.3 $e$ on the atoms of Au. 
The dipoles are allowed  to rotate three-dimensionally
in response to the motion of all atoms in the solvent.
Furthermore, the FFI doubles the number of the Lennard-Jones sites
over the actual number of the Au atoms and locates them between
positions of the real Au atoms in order to achieve a better
agreement with the results obtained by the density functional theory.
Parameters $\sigma_{AuAu}$ and $\epsilon_{AuAu}$ are 0.32 nm and 0.65 kJ/mol for these
interactions.

In all models, the covalent bond between the S atom of cysteine and 
the nearest Au atom on the gold surface needs a special procedure.
Here, this bond is considered to be transient and is
treated in analogy to the method proposed in ref. \cite{nawrocki_2014}
in the context of ZnS. In the case of ZnS, the covalent coupling is through the
disulfide bond which may form between S on the cysteine and S on the surface.
Here, the S-Au bond is also represented by the Morse potential 
(see also Mahaffy $et$ $al.$ \cite{Mahaffy})
\begin{equation}
V_{M}(r_{ij})=D_{ij}[1-exp(-\beta_{ij}(r_{ij}-b_{ij}))]^{2}
\end{equation}
where $i$ and $j$ label the atoms involved. They are in a distance
of $r_{ij}$ and $b_{ij}$ denotes location of the minimum.
Parameter $D_{ij}$ denotes the depth of the potential well,
or -- equivalently -- the dissociation energy.
At small deviations from the minimum,
$V_{M}(r_{ij})\approx \frac{1}{2}k_{ij}(r_{ij}-b_{ij})^2$,
but at large distances $V_M$ disappears. Parameter
$\beta_{ij}$ is related to the effective spring constant $k_{ij}$
through $\beta_{ij}=\sqrt{\frac{k_{ij}}{2D_{ij}}}$.
The bond angle potential between the Au-S-C$_{\beta}$ triplet  and
the dihedral angle potential between the Au-S-C$_{\beta}$-C$_{\alpha}$ quadruplet 
are multiplied by the normalized and sign-inverted Morse potential to
guarantee their decays at large distances. We do not take into account
dissociation of H from the thiol group (S--H) into the solvent 
due to the formation of the S--Au bond.
The functions representing covalent bond between S and Au  
within the framework of FFB
are parametrized according to Jung $et$ $al.$ \cite{jung_1998}, but 
$D_{ij}$ are derived from ref. \cite{lou_2007}. To summarize, we
take $k_{ij}$=82843.2 kJ/mol/nm$^2$, $D_{ij}$=253.6 kJ/mol, and
$b_{ij}$=0.2531 nm. 
For the other two force fields such parameters do not seem to be
available in the literature and the covalent bonding to cysteine
in FFV and FFI will not be discussed here.
S in methionine is considered as not making any covalent bond with gold.

\subsection{Geometry of the System}

Crystalline gold forms the fcc lattice. The unit cell contains 8 atoms
at the corners of a cube and 6 at the center of each face of the cube.
The lattice constant, $a$, {\it i.e.} the  cube edge,
is 0.408 nm long and the distance between the nearest
atoms is equal to 0.288 nm.
We build the solid by repeating the
elementary unit in all Cartesian directions. The most common cleavage
is one that forms the (111) planes \cite{bruce_1988}. 
Atoms on the surface are thus arranged into the hexagonal lattice.

We model the Au surface by a slab of three atom layers located at the 
bottom of the simulation box sized $L_x \times L_y \times L_z$ 
so that the top layer is at $z$=0. Water molecules, ions,
and AAs or peptides are placed  within the box. Here,
$L_x$ and $L_y$ are about 3.5 nm whereas $L_z$ is adjustable. In proper
runs, $L_z$ is equal to 4.0 nm and a reflecting wall is placed at the top.
Above this wall, there is an empty space extending to $z$ = 3 $\times$ $L_z$.
Another reflecting wall, affecting only the AAs, is placed at $z$ = 3.5 nm
-- otherwise an AA might get trapped at the water-vacuum interface.
The purpose of this construction with the empty space is to allow for the usage of 
the periodic boundary conditions (in the periodic image,
above the empty space there are atoms of the solid) combined
with the pseudo-2$D$ particle mesh Ewald summation \cite{essman_1995}.
The solid is considered to be rigid and its geometry is taken to be bulk-like. 
It should be noted that the snapshots of molecules near gold
shown in this paper for illustration pertain to only small fragments
of the simulation box. The size of the fragments can be inferred
by counting the closest distances between the gold atoms.

\subsection{Simulation Settings}

The standard procedure of making the simulations feasible in an acceptable time 
is to apply cutoff radii for all relevant interactions. 
We use the cutoff of 1.2 nm 
combined with the gradual switching the interactions off between 1.0 and 1.2 nm. 
The MD simulations were performed using the leap-frog algorithm 
with the time step of 1 fs. This value of the time step 
is short enough to make the computations stable.
Doubling it may result in generation of prohibitively
large forces at the water-solid interface. 
Temperature control was implemented every 0.1 ps
by using the Berendsen thermostat \cite{spoel_2005} with the temperature of 300 K. 
At the beginning of each run, the energy of the system was minimized 
through the steepest descent algorithm and the initial velocities were Maxwellian. 
The first runs pertained to the situation without any biomolecules -- 
the system was equilibrated for 4 ns with $L_z$ set at 5 nm. 
The duration of equilibration was decided based on the
behavior of the total energy and appeared more than sufficient.
At that stage, $L_z$ was reduced to 4 nm. The reason for making
the reduction in $L_z$  is formation of a depletion layer
in water at the top due to the attraction by the solid at the bottom.
Further equilibration continued for 1 ns and, at that stage,
the biomolecules and Na$^+$ and Cl$^-$ ions were inserted randomly. 
The concentration of the ions corresponded to the physiological 150 mM: 
4 ions of Na$^+$ and 4 ions of Cl$^-$. 
If a biomolecule had a net charge due to the side groups, 
extra ions were inserted to neutralize it. 

The AAs and dipeptides considered in the simulations  are capped by
the acetyl group at the N-terminus and N-methylamide group at the C-terminus. 
These caps eliminate the terminal charges and mimic the presence of a peptide chain 
in which the AA exists in the unionized form (see, $e.g.$ \cite{dragneva_2013}). 
Histidine is considered in its three possible protonation states: 
HIE (H on the $\epsilon$ N atom), HID (H on the $\delta$ N atom) 
and positively charged HIP (H on both $\epsilon$ and $\delta$ N atoms). 
We assume pH of 7. At his value,
all three forms are present with the same probability. 

\subsection{Determination of the Potential of Mean Force}

The main objective of the simulations is to determine the PMF
for twenty AAs, their analogues and dipeptides. The PMF is defined as
an effective potential that corresponds to the average force \cite{kirkwood_1935} 
and it is associated with the center of mass (CM) of the molecules 
(amino acid and dipeptides without the caps). We determine it by
implementing the umbrella sampling method \cite{kumar_1992,lemkul_2010}. 
It involves two stages. In the first stage, one generates a set
of initial conformations for representative values of $z$ by pulling
the CM of the object along the $z$-axis -- perpendicular to the surface of
the solid. The origin of the $z$-axis is at the center of the top atoms of gold.
Pulling is implemented through a "dummy particle" which moves towards the surface 
with a constant speed of 1 nm/ns from $z$=2 nm to $z$=0 and drags the CM by
the harmonic force  corresponding to the spring constant of 5000 kJ/(mol nm$^{2}$). 
We consider 36--39 $z$-placements of the harmonic centers
and generate conformations that are saved every 0.1 ps. The histograms
for each centering are found to be overlapping which signifies sufficient
statistics.
The lateral motion is not constrained so the PMF is averaged laterally. 
Overall, the average forces are calculated based on 
between 100 000 and 150 000 conformations for each z-placement.

In the second stage, these conformations are used for further separate runs 
where the $z$-location of the pulling particle is fixed and the CM
is observed to move within a sampling window of width $\Delta z$. 
The elastic parameter is chosen so that $\Delta z$'s from neighboring
bins overlap.
The distribution of the resulting vertical locations of CM in the window has a
maximum where the harmonic pulling force balances all forces acting  along the $z$ direction. 
This force is averaged over time and locations within each window 
and is then integrated over $z$ to get the PMF.  
The errors in the values of the average
forces are determined by the block averaging method \cite{hess_2002} 
and are propagated during integration. 
Simulations last between 10 ns and 15 ns 
until the convergence of block averages is achieved. 
1 ns corresponds to equilibration. The standard deviations 
of the average $z$ values within the umbrella sampling intervals 
are found to be negligible. 

\subsection{Contact Angle Calculations}

The contact angle of water on the surface of gold
in Figure \ref{droplet} is determined by following
a procedure developed by Ruijter {\it et al.} \cite{ruijter_1999} and later
adopted by Werder {\it et al.} \cite{werder_2002} for water-carbon interaction. 
First, the Cartesian space coordinates of the droplet atoms are transformed 
to the planar coordinates ($r$,$z$), where $r$ is the distance 
from the CM of the droplet to a point on the surface. 
Second, the $r-z$ plane is covered by a fine rectangular grid 
and the time averaged density of atoms is evaluated.
Third, the density profiles in each horizontal layer of grid
points is fitted to a sigmoidal function and the center locations
of the sigmoids are determined (see Fig. \ref{droplet}).
The top and bottom portions of the droplet are discarded in these fits.
Finally, a circle  is fitted to these points  and the contact
angle between the surface and a tangent to the circle is measured. 

\section{Results and Discussion}

\subsection{Properties of Water Near the Gold Surface} 

From now on, we consider simulations in which water molecules
in the starting state are spread out throughout the box, instead 
of forming a droplet.

The time-averaged water density profiles for the  FFV, FFB, and FFI
force fields are shown in Fig. \ref{density.
The bin sizes approximately correspond to the size of one water molecule
(0.142 nm).}
The profile above the hydrophilic FFV surface shows two well
articulated layers with a minimum in the density between them.
This behavior is consistent with refs. \cite{verde_2009,chang_2008}
and the differences in the peak heights are due to different choices
of the bin widths. It is also similar to what one observes
in Lennard-Jones fluids \cite{cieplak_1999,cieplak_2001}.
The region between the first value of $z$ corresponding to a 
non-zero density and the center of the gap is
referred to as the first layer (layer I). 
The second layer (II)
extends to the center of the second (and less articulated) gap.
At still farther distances, we observe the nearly homogeneous
bulk-like behavior. 
In the hydrophobic FFB case, on the other hand, there is a more
gradual transition to the bulk value of the density, though 
regions I and II can still be identified as indicated in
Fig. \ref{bizzarri_den_pol}.
Finally, FFI clearly leads to the hydrophilic FFV-like features
albeit with a more gradual buildup of the first layer on 
moving away from the solid in a qualitative agreement with ref.
\cite{hoefling_2010_a}. Note also that the first layer in FFI is
broader than in FFV and subsequent layers are, therefore, 
further away from the surface by about 0.1 nm.

\begin{figure}[ht]
\begin{center}
\includegraphics[scale=0.27]{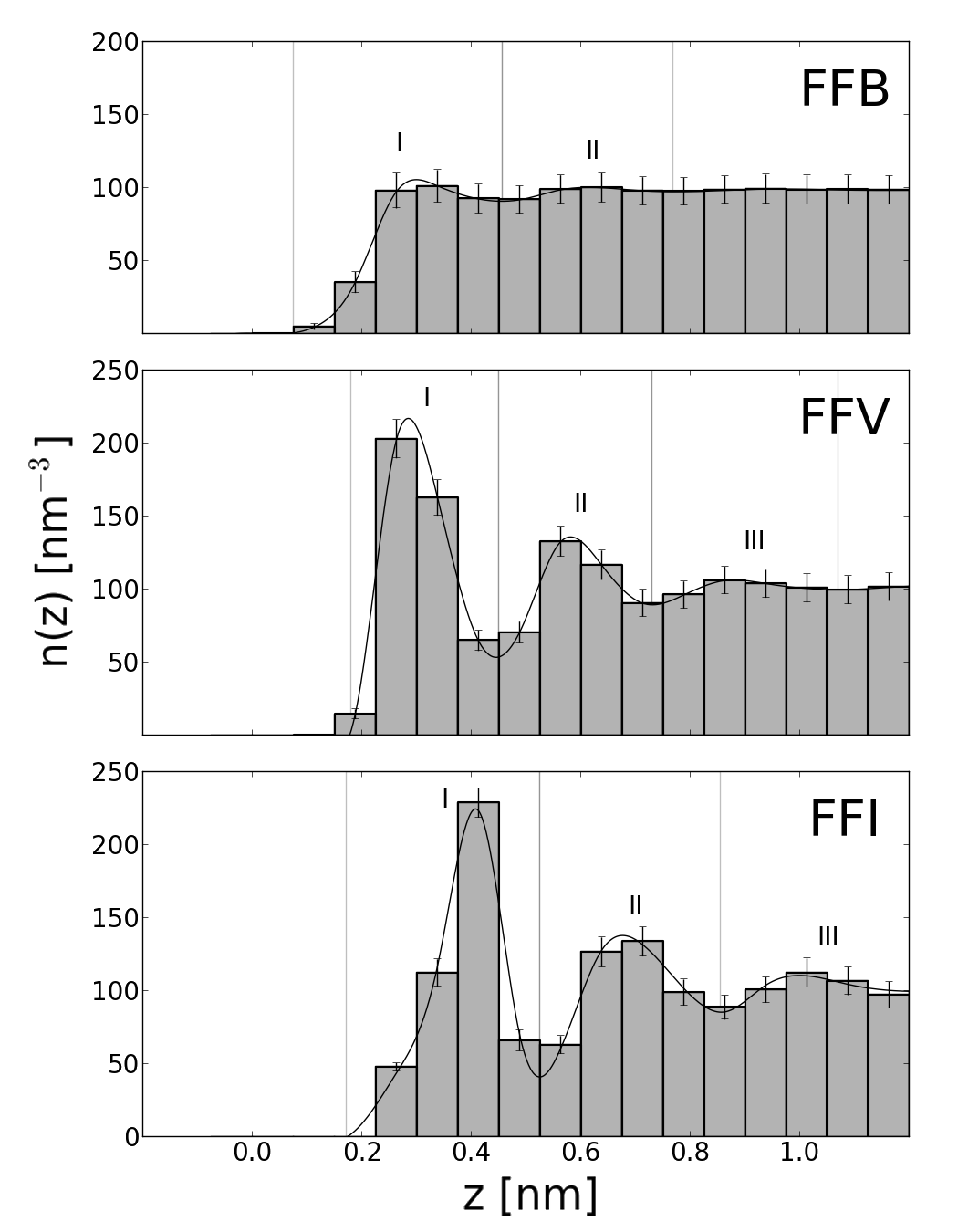}
\caption{ 
   The number density profile of water atoms above the gold surface
   for the three force fields studied here. The $z$
   coordinate is measured from the center of the topmost atoms of the solid.}
\label{density}
\end{center}
\end{figure}

\begin{figure}[ht]
\begin{center}
\includegraphics[scale=0.2]{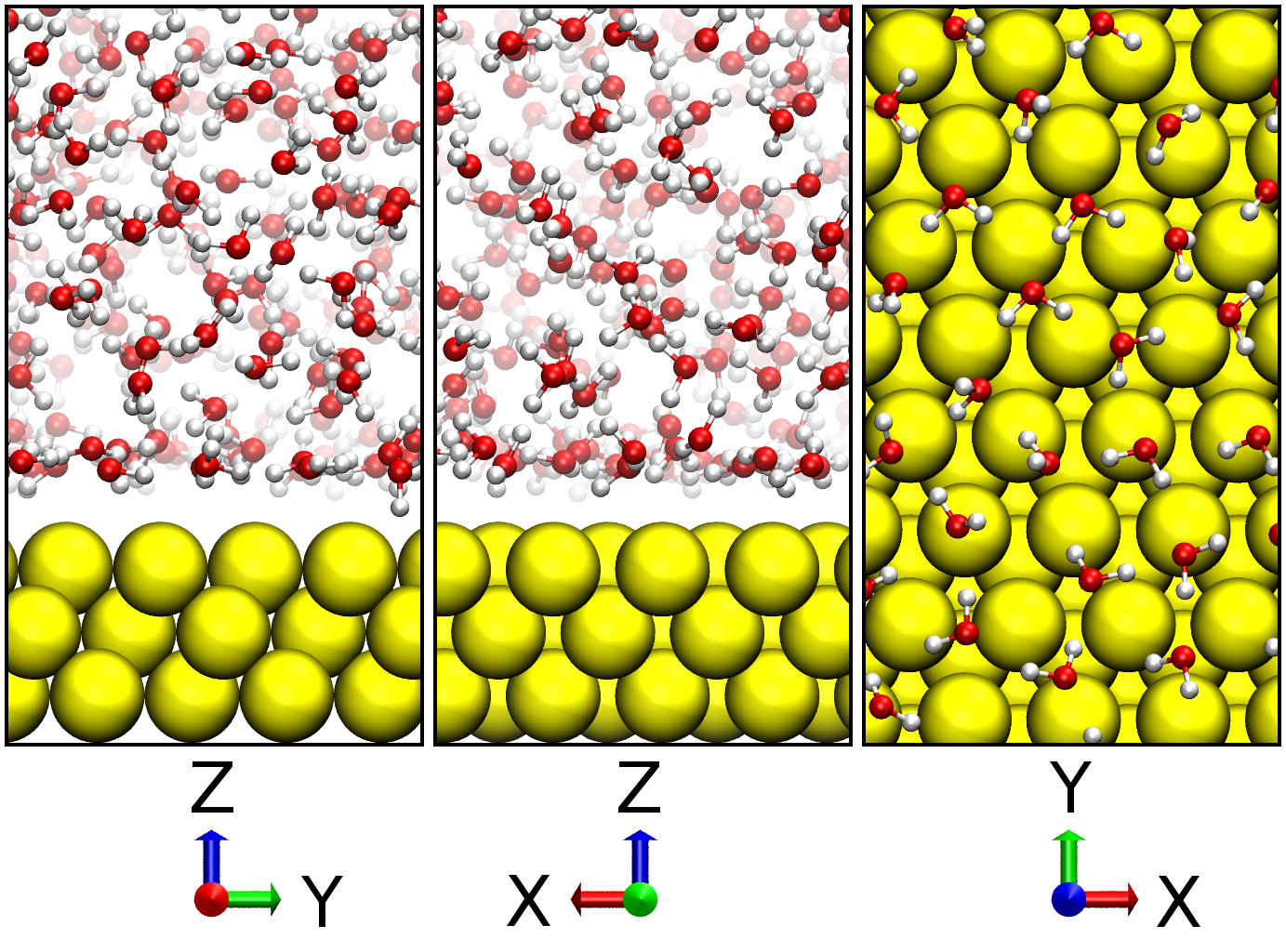}
\end{center}
  \caption{Snapshots of the solid-water interface for the hydrophilic gold surface at various projections. 
    The $z$ direction is perpendicular to the solid.
    The symbols in yellow show atoms of the solid. 
    The remaining symbols show the atoms of water 
    (only those that are close to the solid are shown in the rightmost panel):
    the O atoms are in red and the H atoms in white.}
  \label{verde_sur_sol}
\end{figure}
\begin{figure}[ht]
\begin{center}
\includegraphics[scale=0.2]{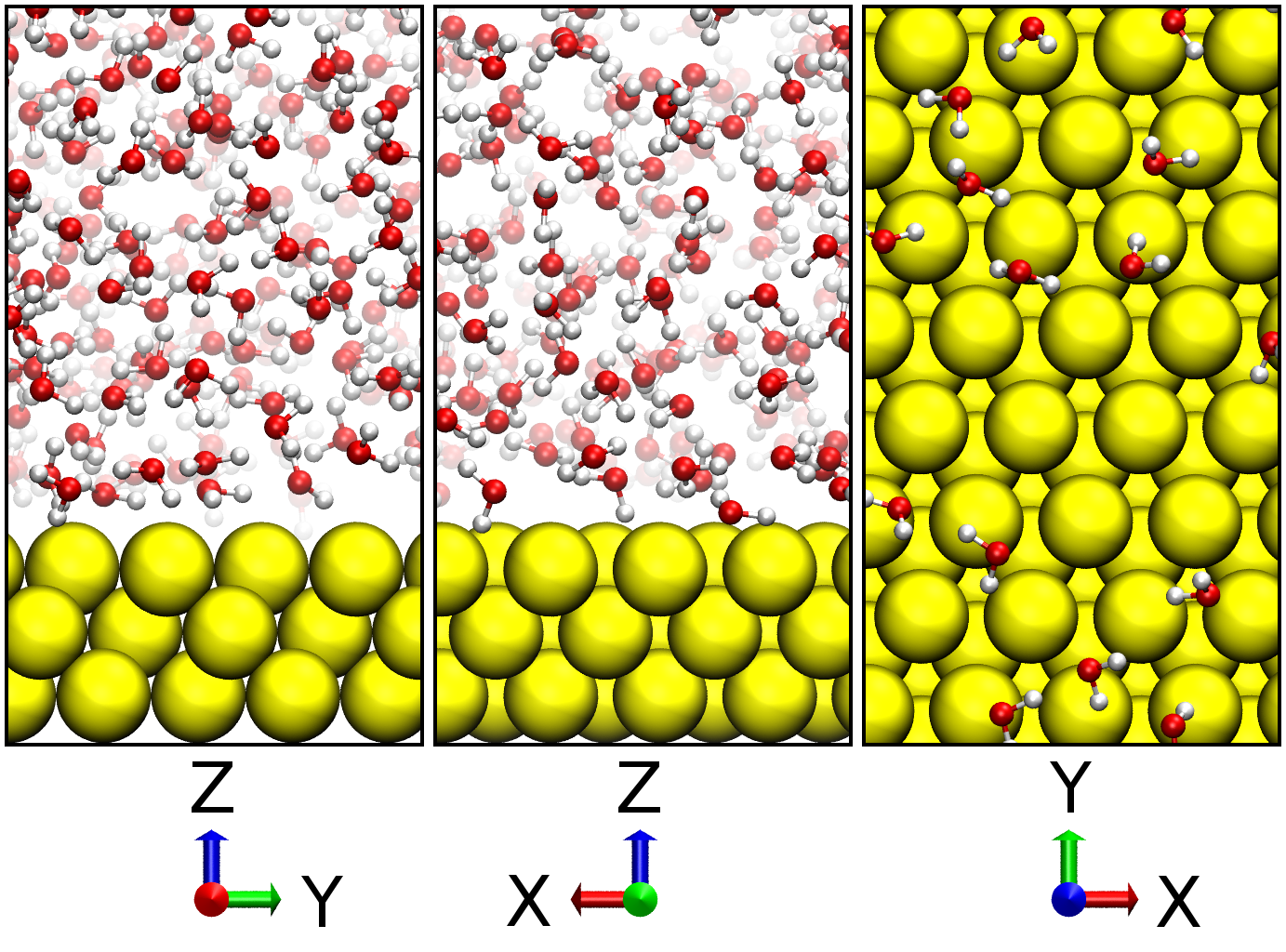}
\end{center}
  \caption{Similar to Fig. \ref{verde_sur_sol} but for the hydrophobic FFB case.}
  \label{bizzarri_sur_sol}
\end{figure}
\begin{figure}[ht]
\begin{center}
\includegraphics[scale=0.2]{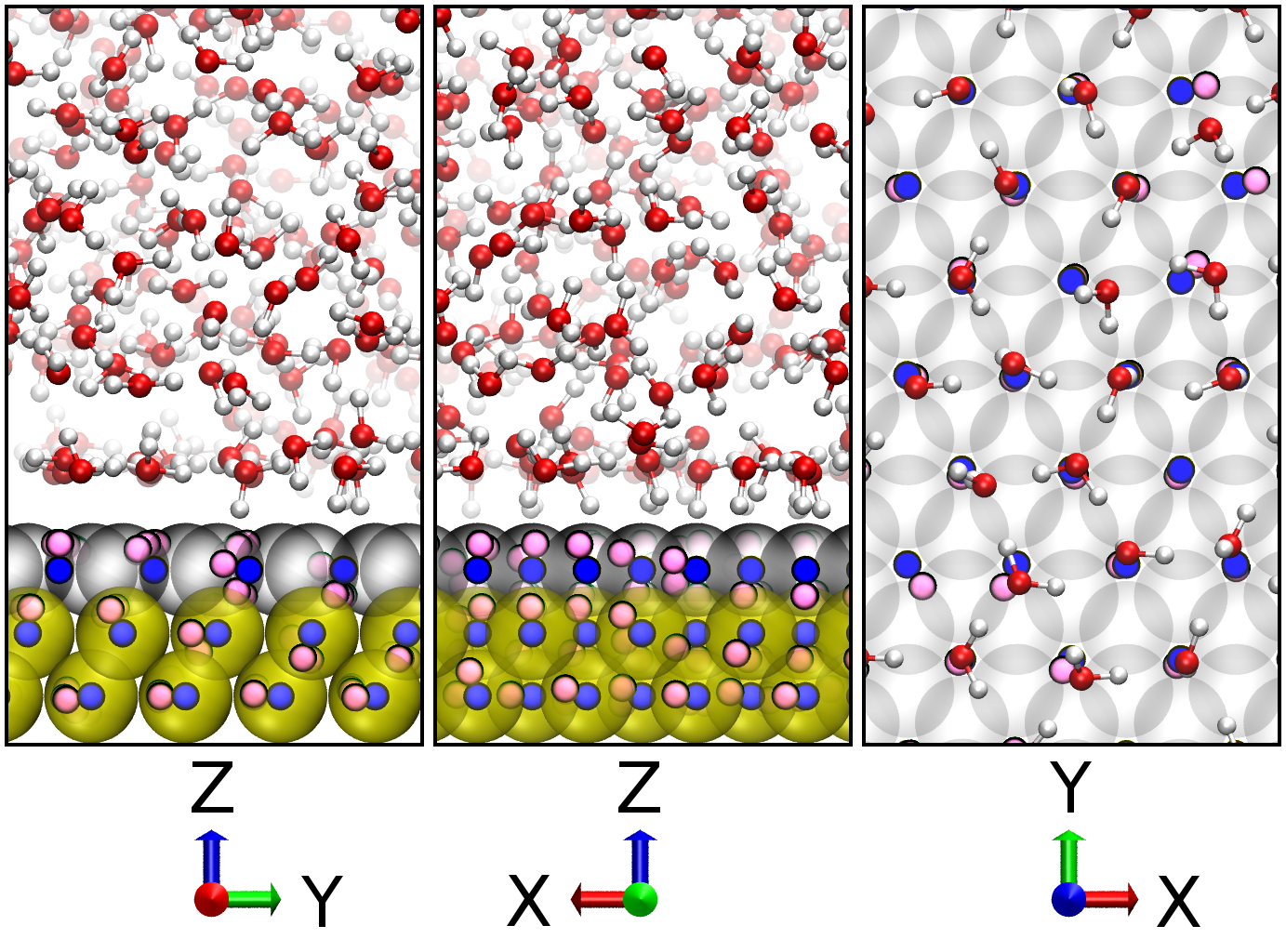}
\end{center}
  \caption{Similar to Fig. \ref{verde_sur_sol} but for the hydrophilic and polarizable FFI case.
  The additional graphical symbols pertain to the description of the polarization
   of gold. The gray spheres correspond to the additional Lennard-Jones centers
   which are placed at the top layer of gold.
   The blue-mauve doublet (of spheres with the coat) corresponds to the dipole associated with the Au atom. 
   The distance between the local charges is equal to 0.07 nm.}
  \label{iori_sur_sol}
\end{figure}

The behavior of water molecules in layer I can also be glimpsed
from the snapshots in Figs. \ref{verde_sur_sol}, \ref{bizzarri_sur_sol},
and \ref{iori_sur_sol} for FFV, FFB, and FFI respectively.
For FFB, there are much fewer molecules near the surface than
for FFV, but they come closer to the surface. For FFI, the
situtation is between the two. 

\begin{figure}[ht]
\begin{center}
\includegraphics[scale=0.27]{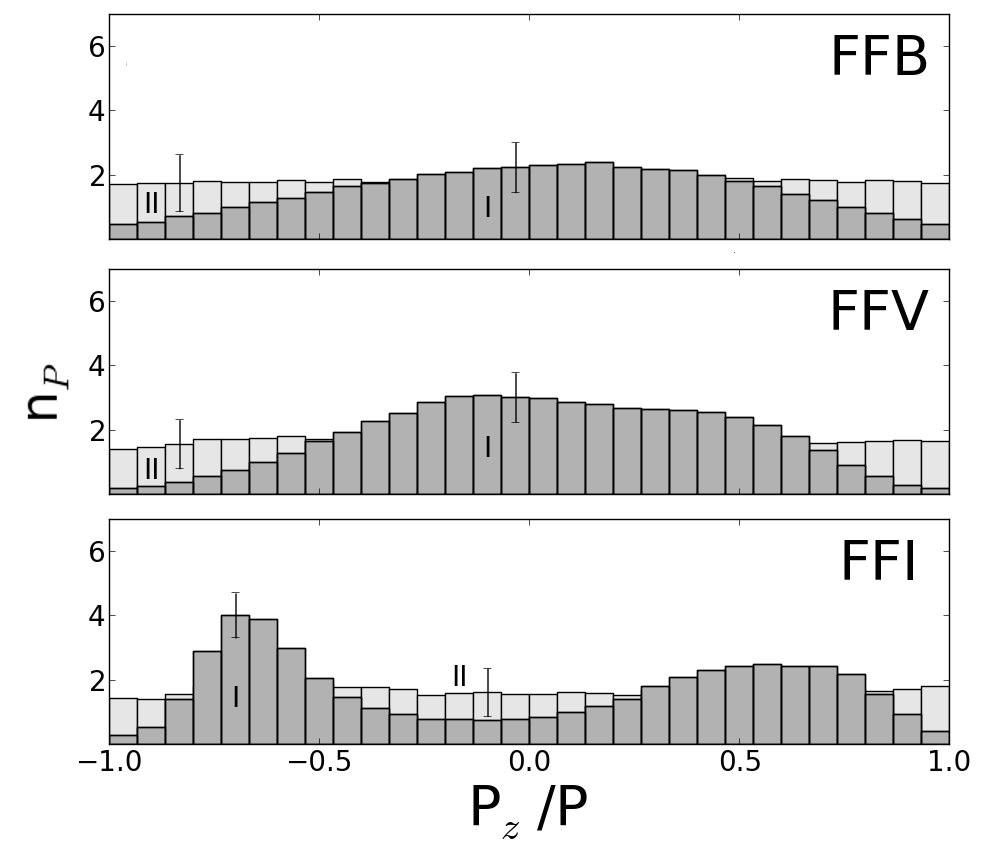}
\caption{The distributions of water polarization for the first 
          two water layers
           - the darker gray color corresponds to the first layer
           and the lighter to the second. $P$ is the magnitude of the polarization
           vector and $P_{z}$ denotes its $z$-component.
           $n_P$ is the number of water molecules with a given polarization
           divided by the number of all water molecules in the two layers
           and expressed as a percentage.
           The contents of the bins in the two layers add up together to 100\%.
           The average errors for both layers are shown in selected bins.}
\label{polari}
\end{center}
\end{figure}

Fig. \ref{polari} 
shows the distributions of the $z$ component of the electric 
polarization of water in layers denoted 
as I and II in the corresponding panels at the top. 
These plots for FFV and FFB 
point to a substantial orientational disorder, even bigger than in the 
case of the ZnO \cite{nawrocki_2013} and ZnS -- see Fig. 2 in ref.\cite{nawrocki_2014}.
The case of ZnO is shown in the Supplementary Information (SI).
Since for FFV and FFB the molecules of water are observed to
prefer orientations that are parallel to the surface,
the distributions of P$_{z}$/P are peaked close to P$_z$/P=0.
There is some more electric order in the hydrophilic case and the molecules
are seen to prefer alignments along the $x$- and $y$-directions 
(not shown). 
The case of FFI, however, is distinct: the molecules in layer I prefer
nearly vertical local polarizations with a half of the molecules
pointing up and the other half -- down. This feature contributes
to the broader width of layer I for FFI and distinguishes between
the two hydrophilic models. In FFI, the molecules of water
tend to be located just above the dipoles representing the
polarization of gold.

\subsection{Potentials of the Mean Force for Amino Acids} 

\begin{figure}[ht]
\begin{center}
\includegraphics[scale=0.4]{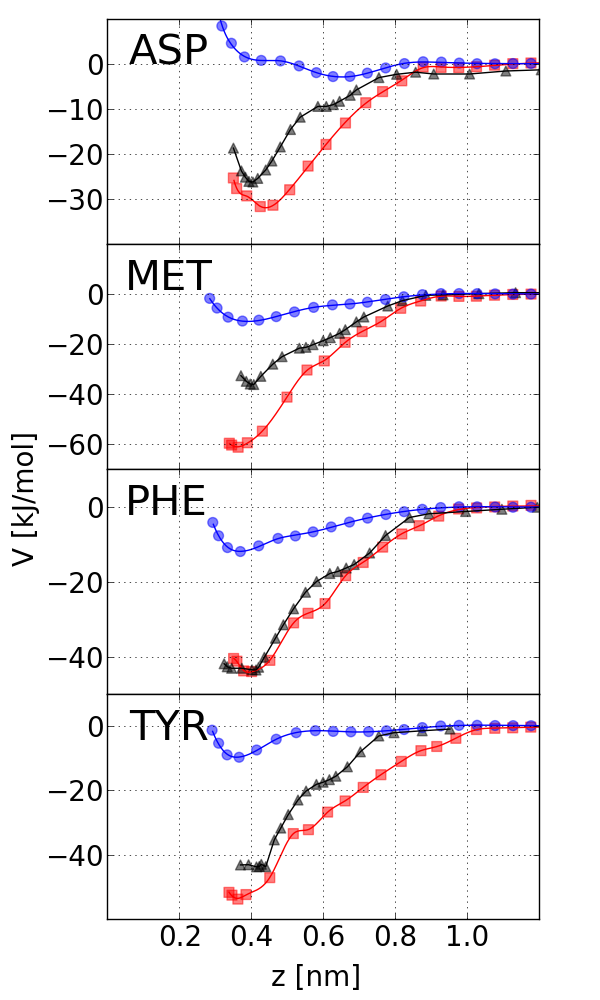}
\caption{The plots of PMFs as a function of $z$ for four capped amino acids within
the three force fields considered. The lines correspond, top to bottom, to
FFB (circles, blue line), FFI (triangles, black) and FFV (squares, red). 
The data points for the FFI case come from
ref. \cite{hoefling_2010_b}. Since the authors do not provide the real distances
from the surface, we shift the position of the central line so that
the minimum is set at 0.4 nm.}
\label{pmf_com}
\end{center}
\end{figure}

Fig. \ref{pmf_com} shows PMFs for  the capped ALA, MET, PHE, and TYR for
the three force fields considered here. The hydrophobic FFB leads to the potential
wells which are shallow and FFV to wells which are deep. FFI is in between,
except that for PHE it almost coincides with the line for FFV.
Figs. 5 and  6 in SI show  plots of the PMFs
for all twenty capped AAs near the FFV
and FFB surfaces respectively.
Most potentials have a single and well defined minimum, but there are
also some, especially in the FFB case (see ASN, GLN, LYS, ARG),
which come with two minima.
The binding energies and bond lengths
($i.e.$ the locations of the minima in the PMF) corresponding
to the lowest minimum
are summarized in Tab. \ref{bizzarri_verde_aa_tab}. 
For FFB, the error bars vary between 1.0 and 2.5 kJ/mol.
For FFV, they are larger in absolute terms as they vary between 3.1 and 13.6 kJ/mol
but smaller percentage wise. The error bars in the parameter sigma are
always smaller than 0.0022 nm and are not listed.
The table also shows
results for FFI obtained by Hoefling {\it et al.} \cite{hoefling_2010_b}.
Our own recalculation for PHE confirms their value within the error bars.
Unlike what happens for ZnO and ZnS, we find no AA that would be 
repelled by gold. 

\begin{table}[ht]
\begin{tabular}{cccccc}
\multirow{2}{*}{AA} & \multicolumn{2}{c}{hydrophobic FFB} &
\multicolumn{2}{c}{hydrophilic FFV} & polarizable FFI \\
\cline{2-6}
& $\epsilon$ & $\sigma$ & $\epsilon$ & $\sigma$ & $\epsilon$ \\
& [kJ/mol] & [nm] & [kJ/mol] & [nm] & [kJ/mol] \\
\hline
ASP  &    2.7 $\pm$ 1.2  &  0.65  &   31.7 $\pm$ 4.9  &  0.44  & 25.5 \\
GLU  &    2.2 $\pm$ 1.2  &  0.66  &   47.9 $\pm$ 7.9  &  0.43  & 17.5 \\
CYS  &    7.3 $\pm$ 1.4  &  0.38  &   39.7 $\pm$ 5.2  &  0.37  & 37.7 \\
\textcolor{gray}{mCYS} & \textcolor{gray}{251.1 $\pm$ 2.5} & \textcolor{gray}{0.36} &   --  & --  & -- \\
ASN  &    3.6 $\pm$ 1.3  &  0.38  &   31.7 $\pm$ 6.7  &  0.41  & 26.1 \\
PHE  &   11.7 $\pm$ 1.9  &  0.37  &   43.9 $\pm$ 4.9  &  0.39  & 43.6 \textcolor{gray}{(35.9$\pm$9.8)} \\
THR  &    4.8 $\pm$ 1.4  &  0.49  &   35.2 $\pm$ 7.3  &  0.39  & 28.9 \\
TYR  &    9.4 $\pm$ 2.2  &  0.36  &   53.3 $\pm$ 17.7  &  0.36  & 44.2 \\
GLN  &    5.1 $\pm$ 1.8  &  0.40  &   56.6 $\pm$ 11.2  &  0.36  & 28.6 \\
SER  &    4.3 $\pm$ 1.3  &  0.40  &   30.7 $\pm$ 3.1  &  0.41  & 23.1 \\
MET  &   10.6 $\pm$ 1.8  &  0.39  &   60.7 $\pm$ 12.1  &  0.36  & 39.3 \\
TRP  &   12.9 $\pm$ 2.3  &  0.34  &   58.6 $\pm$ 11.5  &  0.34  & 40.2 \\
VAL  &    9.3 $\pm$ 1.2  &  0.43  &   38.4 $\pm$ 8.0  &  0.40  & 24.8 \\
LEU  &   11.7 $\pm$ 1.4  &  0.44  &   34.6 $\pm$ 6.7  &  0.41  & 25.4 \\
ILE  &   10.9 $\pm$ 1.5  &  0.42  &   27.2 $\pm$ 8.8  &  0.50  & 25.1 \\
GLY  &    4.3 $\pm$ 1.2  &  0.36  &   36.9 $\pm$ 7.6  &  0.33  & 23.6 \\
ALA  &    6.5 $\pm$ 1.2  &  0.40  &   31.0 $\pm$ 3.9  &  0.37  & 21.9 \\
PRO  &    8.1 $\pm$ 1.2  &  0.41  &   37.0 $\pm$ 7.2  &  0.36  & 26.0 \\
\textcolor{gray}{HIE} & \textcolor{gray}{3.4 $\pm$ 1.9} & \textcolor{gray}{0.38} &
\textcolor{gray}{52.8 $\pm$ 11.0} & \textcolor{gray}{0.38} & \multirow{2}{*}{34.0} \\
\textcolor{gray}{HID} & \textcolor{gray}{2.9 $\pm$ 1.1} & \textcolor{gray}{0.63} &
\textcolor{gray}{50.2 $\pm$ 13.6} & \textcolor{gray}{0.35} & \\
\textcolor{gray}{HIP} & \textcolor{gray}{0.9 $\pm$ 1.2} & \textcolor{gray}{0.66} &
\textcolor{gray}{41.8 $\pm$ 9.5} & \textcolor{gray}{0.36} & $\sim$40.8 \\
LYS  &    3.8 $\pm$ 1.4  &  0.67  &   40.7 $\pm$ 7.7  &  0.38  & 30.0 \\
ARG  &    5.8 $\pm$ 2.5  &  0.36  &   80.3 $\pm$ 12.9  &  0.35  & 36.3 \\
\hline
AAs\textsuperscript{\emph{a}} & \textbf{6.9$\pm$3.4} & \textbf{0.44$\pm$0.10} &
\textbf{43.2$\pm$12.9} & \textbf{0.39$\pm$0.04} & \textbf{30.1$\pm$7.5} \\
AAs\textsuperscript{\emph{b}} & \textbf{18.5$\pm$52.1} & \textbf{0.44$\pm$0.10} & -- & --
& -- \\
\hline
\end{tabular}
\\
\textsuperscript{\emph{a}} Average and dispersion for all AAs with CYS and
\textsuperscript{\emph{b}} with mCYS.
\caption{Values of the binding energy, $\epsilon$, and of the bond length, $\sigma$, 
between the CM of an AA and the surface of gold
for the three force fields discussed in the paper.
The values for FFI are after
Hoefling {\it et al.} \cite{hoefling_2010_b} (this reference does not
reveal which form of the uncharged HIS has been used; we have read off 
the value for HIP from the plot of the PMF).
Our own simulation with FFI
for PHE is indicated in gray.
CYS denotes cysteine that does not form the covalent bond
and mCYS one that does. Some values were not determined as indicated
by the hyphen.
The average and dispersion of $\epsilon$ and $\sigma$ that are listed at the bottom. 
Results corresponding to the various forms of histidine are first averaged to form
one entry when determining the overall averages.}
\label{bizzarri_verde_aa_tab}
\end{table}

All AAs are observed to have optimal binding resulting from a
direct interaction with gold (only spot-checked for FFI, but
fully for FFV and FFB) as reflected in the 
values of $\sigma$ in Table \ref{bizzarri_verde_aa_tab}.
This is distinct from the situation with ZnO and ZnS where water 
affects the couplings significantly. In the case of ZnO \cite{nawrocki_2013}
adsorption of AAs occurs through the first layer of water (the CM 
of an AA is above it). For ZnS \cite{nawrocki_2014}, this is the
situation for most of the AAs.

On average, FFV yields $\epsilon$ that is about
six times stronger than FFB, and FFI -- about four times stronger.
The specificity can be measured by the ratio of the standard
deviation to the mean over the AAs. It is 49\%, 30\% and 25\%
for FFB, FFV, and FFI respectively. Thus the hydrophobic surface
yields smaller couplings but distinguishes between AAs better.
FFI is the least specific.

The strongest binders in FFB are the hydrophobic TRP, LEU and PHE 
because binding with the hydrophobic surface minimizes their contact
with water.
In FFV the strongest binders are the charged ARG, and hydrophobic MET 
and TRP, then polar GLN and 
the uncharged forms of HIS (HID and HIE). 
This is probably due to a good spatial adjustment of the 
atoms in large residues to the hexagonal network of the gold atoms.
For FFI, the aromatic TYR, PHE, and TRP bind the best. 
The FFV results are consistent with the preferences established
experimentally, as summarized by Corni {\it et al.} \cite{corni_2013}.
Also, the fact that the adsorption energy of ARG is twice as big
as for LYS, despite the fact that both are positively charged,
is consistent with the experimental results obtained 
for polyarginine and polylysine \cite{willett_2005}.

\begin{figure}[ht]
\begin{center}
\includegraphics[scale=0.4]{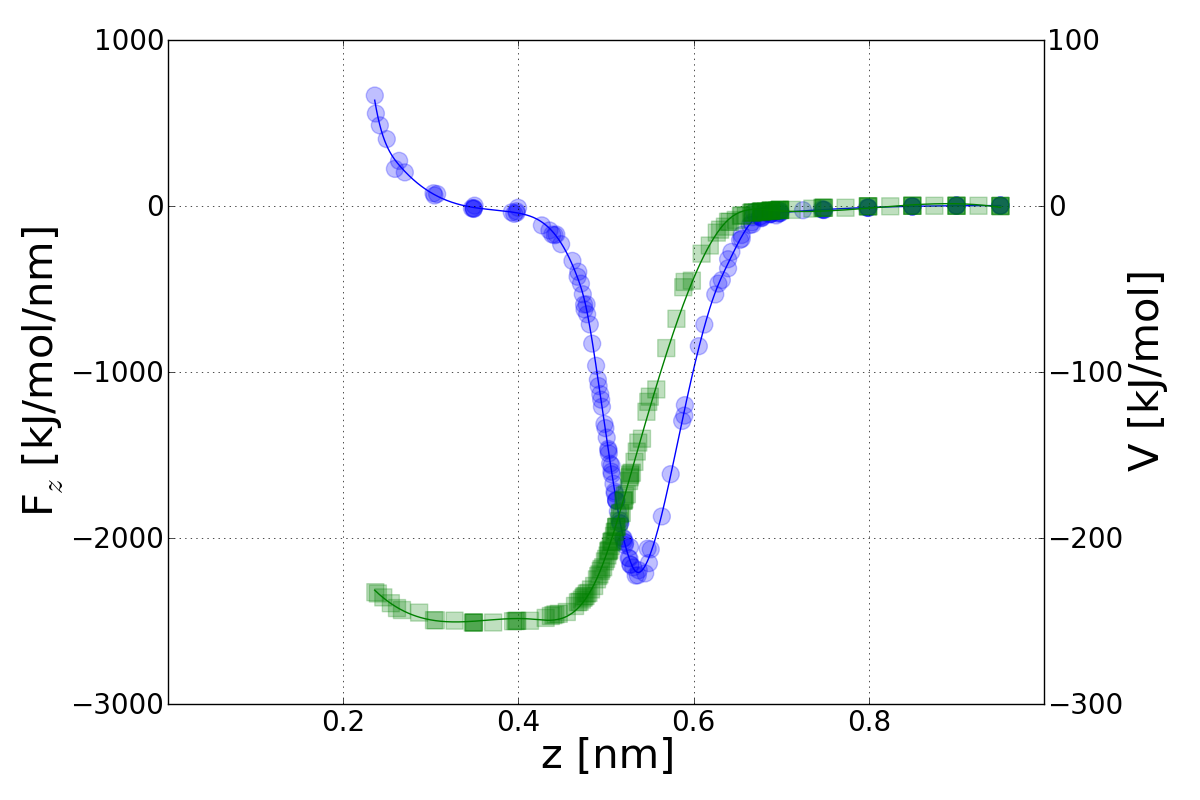}
\caption{The force on mCYS in water solution (blue circles, the left vertical 
axis) and the corresponding PMF (green squares, the right vertical axis) as
derived with the FFB force field and as a function of $z$.
The covalent bond is modeled by the Morse potential. }
\label{bizzarri_cysau_pmf}
\end{center}
\end{figure}

Table \ref{bizzarri_verde_aa_tab} has two lines that relate to cysteine.
One is denoted as CYS and another by mCYS. The former does not
include the covalent bond and the latter does. 
The notation is as in ref. \cite{nawrocki_2014} and "m" stands for
"modified". It is easier to include the covalent
bond in the hydrophobic FFB case, since Bizzarri {\it et al.}
\cite{bizzarri_2003b} have provided parameters for its description
by the permanent harmonic coupling. We replace the harmonic coupling by
the Morse potential and make use of the corresponding harmonic parameters.
Determination of the PMF in this case required more frequent distance 
sampling of the attractive region 
and application of a wide range of the pulling spring constants 
in the umbrella sampling.
The resulting average force and PMF are shown in Fig. \ref{bizzarri_cysau_pmf}. 
The binding energy of 251.1 kJ/mol is 34 times larger that
when the covalent effects are not included and is about the same as 
the dissociation energy of S-Au bond (253.6 kJ/mol) \cite{lou_2007}. 
A similar enhancement is expected for the other two force fields.
It is interesting to note that even though the potential well gets
deeper, its width remains about the same. Thus the protein must
get to the surface to form a bond.
The migration at distances
larger than 1 nm from the surface is through difussion
and at distances that are shorter -- through a combination of diffusion and the
pull of the surface modified by the adsorbed water.
We find that a free evolution of cysteins ($i.e.$ not constrained
as in the calculation of the PMF) leads to a rapid adsorption to gold
(within 10 ns) and to no desorption events during the time scale
of the simulations -- eight trajectories of 20 ns each. 

\subsection{Amino Acids Side Chains Analogues}

A capped AA may bind to the surface through various groups:
C$^\alpha$-H, -NH and -CO in the peptide bond, side groups
or their fragments, as well as the caps CH$_{3}$-HN- and -CO-CH$_{3}$. 
In order to bring the role of the side groups out
it is interesting to consider 
molecules which are analogues of the side chains. This approach has 
been suggested by the studies of the analogues near the surfaces of
TiO$_{2}$ \cite{monti_2010} and quartz \cite{wright_2012}.
Table \ref{verde_aa_tab} summarizes the results obtained with FFV.
The error bars vary between
1.0 and 2.9 kJ/mol.
The plots of the PMFs are shown in Fig. 7 in the SI.
The highest binding energy in the table is for guanidinium
which is an analogue of a part of the side chain in ARG 
(the side chain consists of the 3-carbon aliphatic chain
which is ended by guanidinium).
This result is consistent with the highest $\epsilon$ found for ARG 
itself (see Table \ref{bizzarri_verde_aa_tab}). The binding energy
for guanidinium is seen to be 35\% of that for ARG. 
 
\begin{table}[ht]
\centering
\begin{tabular}{cccccc}
AN & $\epsilon$ & $\sigma$ & AA & $\epsilon$ & $\sigma$ \\
& [kJ/mol] & [nm] & & [kJ/mol] & [nm]\\
\hline
METHANE  &    4.3 $\pm$ 1.0  &  0.34  &  ALA &   31.0 $\pm$ 3.9  &  0.4 \\
AMMONIUM  &    1.1 $\pm$ 1.0  &  0.59  &  LYS &   \multirow{2}{*}{40.7 $\pm$ 7.7}  &  \multirow{2}{*}{0.4} \\
BUTYLAMMONIUM  &   11.1 $\pm$ 2.9  &  0.39  &  LYS & & \\
METHANOL  &    5.1 $\pm$ 1.3  &  0.35  &  SER &   30.7 $\pm$ 3.1  &  0.4 \\
METHANOATE  &    4.0 $\pm$ 2.9  &  0.43  &  ASP &  \multirow{2}{*}{31.7 $\pm$ 4.9}  &  \multirow{2}{*}{0.4} \\
ETHANOATE  &    6.7 $\pm$ 1.9  &  0.43  &  ASP & & \\
GUANIDINIUM  &   28.2 $\pm$ 1.7  &  0.32  &  ARG &   80.3 $\pm$ 12.9  & 0.4 \\
BENZENE  &   19.4 $\pm$ 1.9  &  0.33  &  PHE &   43.9 $\pm$ 4.9  &  0.4 \\
\textcolor{gray}{WATER}  &    \textcolor{gray}{2.8 $\pm$ 1.0}  &  \textcolor{gray}{0.32}  \\
\end{tabular}
\caption{Values of $\epsilon$ and  $\sigma$ 
for the analogues of the AA side chains and for the molecule of water.
The values are derived from the PMF as calculated with the hydrophilic FFV model.
The last column provides data for the corresponding AAs.}
\label{verde_aa_tab}
\end{table}

The second biggest $\epsilon$ in Table \ref{verde_aa_tab} is for
benzene which is an analogue of the ring in the main 
part of PHE side chain and it constitutes 44\% of the binding strength for PHE.
The third biggest is for butylammonium -- an analogue of 
the complete side chain of LYS (27\%). 
ARG binds about twice as strong than LYS and PHE in FFV
whereas guanidinium 3 times as strong that butylammonium
and 1.5 as strong than benzene.
Generally, the relative $\epsilon$-based ranking of the analogues
(guanidinium, benzene, butylammonium, ethanoate, and methane)
is mostly the same as their corresponding AAs (ARG, PHE, LYS, ASP, and ALA).
Ammonium and metanoate are less complete analogues of their
corresponding AAs and they do not follow the trend.
Despite the exceptions, the existence of the trend suggests that
all groups in capped amino acids, but the side chains,  
have the same contribution to the total binding energy
and the differentiation comes solely from the side chains.

\begin{figure}[h!]
\centering
\includegraphics[scale=0.35]{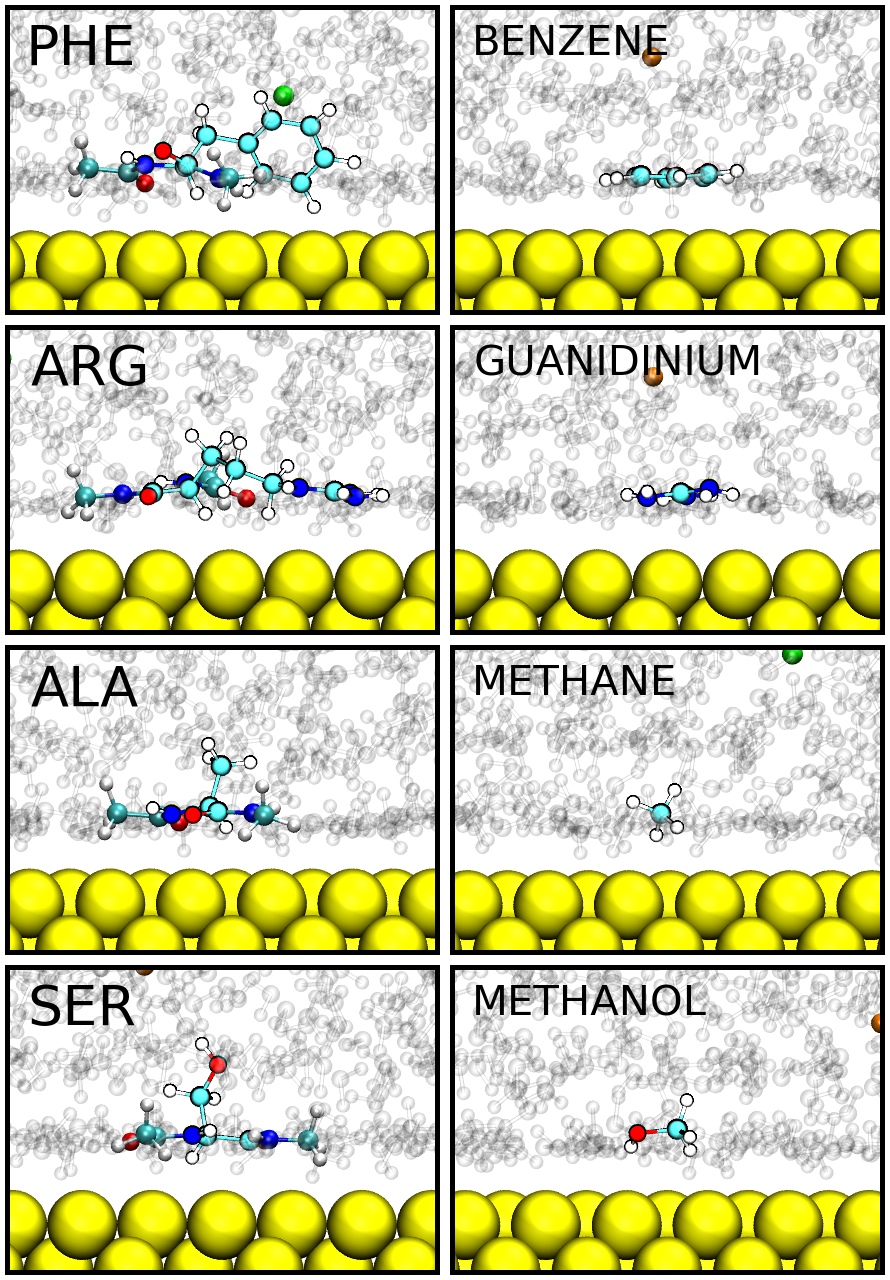}
\caption{Comparison of the optimal conformations of selected AAs and of the
analogues of their side chains (or parts of their side chains) in the FFV model.}
\label{verde_an_pic_b}
\end{figure}

\begin{figure}[h!]
\centering
\includegraphics[scale=0.35]{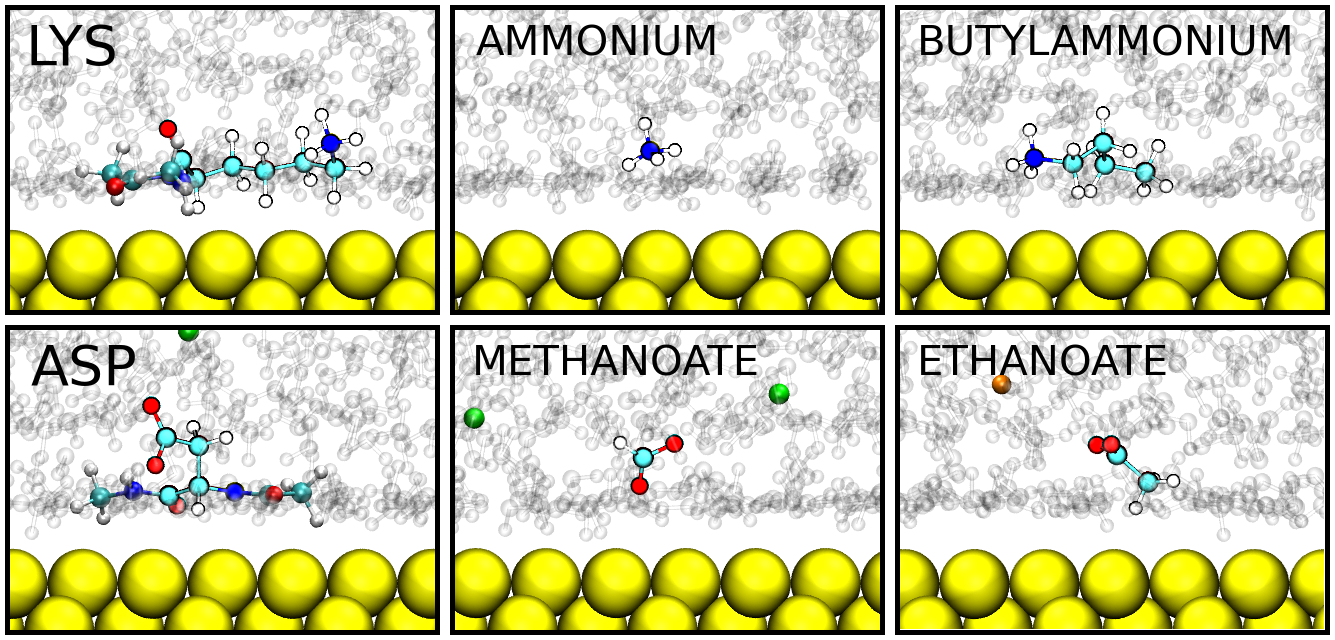}
\caption{Continuation of Fig. \ref{verde_an_pic_b}.}
\label{verde_an_pic_a}
\end{figure}

Examples of the optimal conformations ($i.e.$ those corresponding
to the vicinity of the lowest minimum in PMF) of the analogues shown in 
Figs. \ref{verde_an_pic_b} and \ref{verde_an_pic_a} can be distinct from
their corresponding side chains in the capped AAs. 
Three (ALA, SER, ASP) out of the six AAs shown there
do not bind  to the surface by the side chains but by all other
groups, since the side chains seek water.
This is consistent with the fact that the binding energies 
of the three AAs are just slightly smaller than
that for GLY (36.92 KJ/mol) with H as its side group.
At the same time, their analogues  (methane, methanol, methanoate,
and ethanoate) do bind to the surface.
On the other hand, the remaining three AAs (PHE, ARG, LYS) connect by all
groups including the side chains  and behave like their analogs
which explains the top values of $\epsilon$ for these molecules.
PHE, ARG, and LYS have long side chains and can adjust to the surface better
than the short chains of ALA, SER, ASP, and their analogues.

Notice that only ammonium (in LYS, ammonium and butylammonium) 
does not overcome the first water layer during adsorption. 
This seems to be in agreement with the simulation of single water molecule, 
since ammonium is the only analogue of lower binding energy. 

\subsection{Dipeptides on the Hydrophobic Surface}

We now consider capped dipeptides. Due to the interactions between the
two side groups and the expected relative decrease of the role of the
capped groups, dipeptides should have PMFs
which are not simply superpositions of the potentials
derived for the constituent single AAs (see also ref. \cite{dragneva_2013}).
We shall illustrate this point  here for the hydrophobic FFB model.

First, we select six characteristic AAs:
the positively charged LYS, negatively charged ASP, polar SER, hydrophobic LEU, 
cyclic PHE, and the smallest GLY. We combine them into dipeptides
in all possible ways. Additionally, for one combination, SER--GLY, 
we also consider its reverse, GLY--SER and observe the effect of the
reversal to be minor (the sequence
in a peptide chain is counted from the N to the C termini).

\begin{table}[ht]
\begin{tabular}{cccccccc}
\multicolumn{2}{c}{\multirow{2}{*}{\textbf{$\epsilon$ [kJ/mol]}}} & LYS   & ASP  & SER   & LEU   & PHE   & GLY  \\
\cline{3-8}
\multicolumn{2}{l}{} & 3.8 & 2.7 & 4.3 & 11.7 & 11.7 & 4.3 \\
\hline
LYS & 3.8 & 2.1 $\pm$ 1.2  &      &       &       &       &      \\
ASP & 2.7 & 4.2 $\pm$ 2.0  & --   &       &       &       &      \\
SER & 4.3 & 3.4 $\pm$ 2.0  & 2.0 $\pm$ 1.6 & 2.7 $\pm$ 1.2  &       &       & \textcolor{gray}{3.45 $\pm$ 1.4} \\
LEU & 11.7 & 7.9 $\pm$ 1.8  & 7.4 $\pm$ 2.1 & 9.8 $\pm$ 1.7  & 14.1 $\pm$ 1.9 &       &      \\
PHE & 11.7 & 11.7 $\pm$ 3.0 & 8.8 $\pm$ 2.2 & 11.6 $\pm$ 2.2 & 13.3 $\pm$ 2.2 & 15.7 $\pm$ 3.7 &      \\
GLY & 4.3 & 3.1 $\pm$ 1.6  & 1.3 $\pm$ 1.3 & \textcolor{gray}{3.6 $\pm$ 1.8}  & 10.6 $\pm$ 1.6 & 7.9 $\pm$ 2.1  & 3.7 $\pm$ 1.4 \\
\end{tabular}
\caption{Values of the binding energy
between the CM of a dipeptide and the Au surface 
as determined through the umbrella sampling method within the FFB model.
The error bars vary between 1.2 and 3.7 kJ/mol.
The AA which is the first sequentially ($i.e.$ at the N-terminus)
is listed at the top line,
together with the corresponding single-AA value of $\epsilon$.
The second AA (at the C-terminus) is listed vertically. The entry in gray is for 
SER--GLY which is reverse  to GLY--SER.
Symbol -- signifies a non-binding situation.}
\label{bizzarri_dip_tab}
\end{table}

\begin{figure}[h!]
\centering
\subfigure[]{
\includegraphics[scale=0.4]{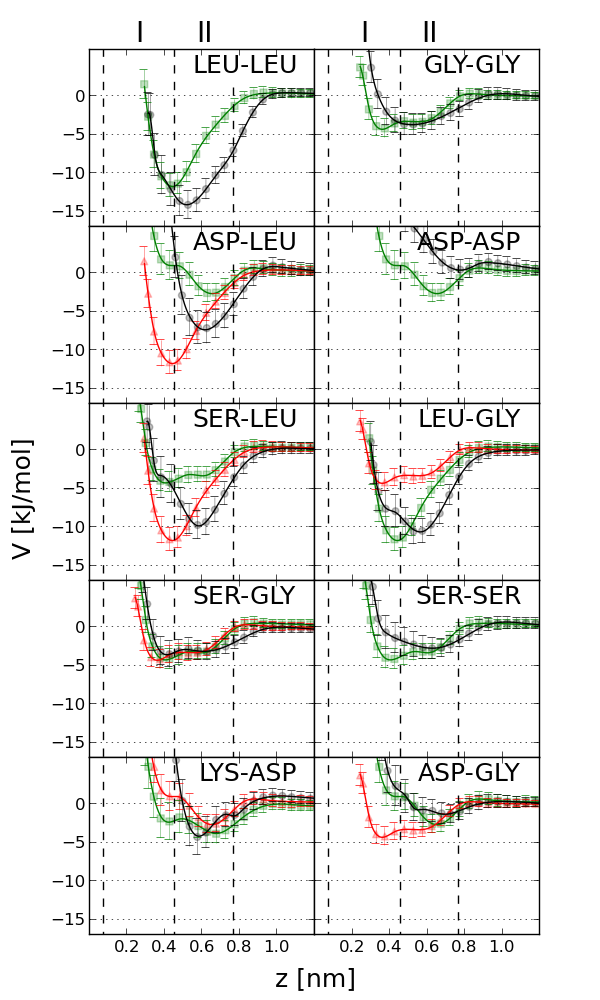}}
\subfigure[]{
\includegraphics[scale=0.4]{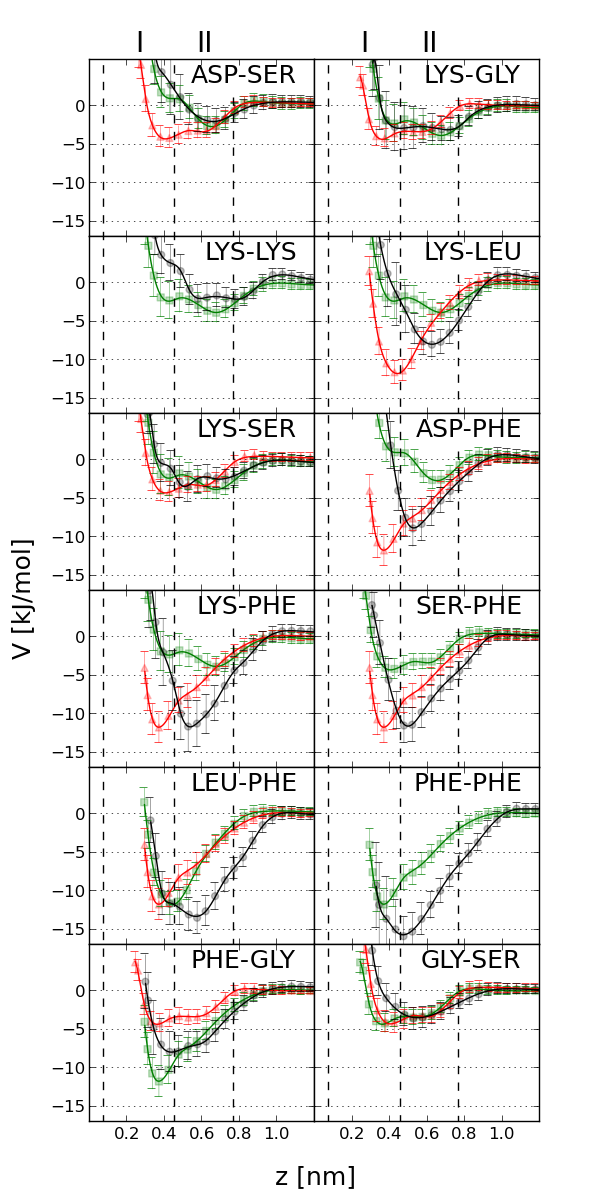}}
\caption{The black circles show the PMF for the indicated capped dipeptides 
        in water solution. 
         The green squares and red triangles show the single AA results: green for the
         first and red for the second AA in the sequence.
         The vertical dashed lines represent the boundaries of the first (I)
         and second (II) water layers.}
\label{bizzarri_dip_pmf}
\end{figure}

\begin{figure}[ht]
\begin{center}
\includegraphics[scale=0.35]{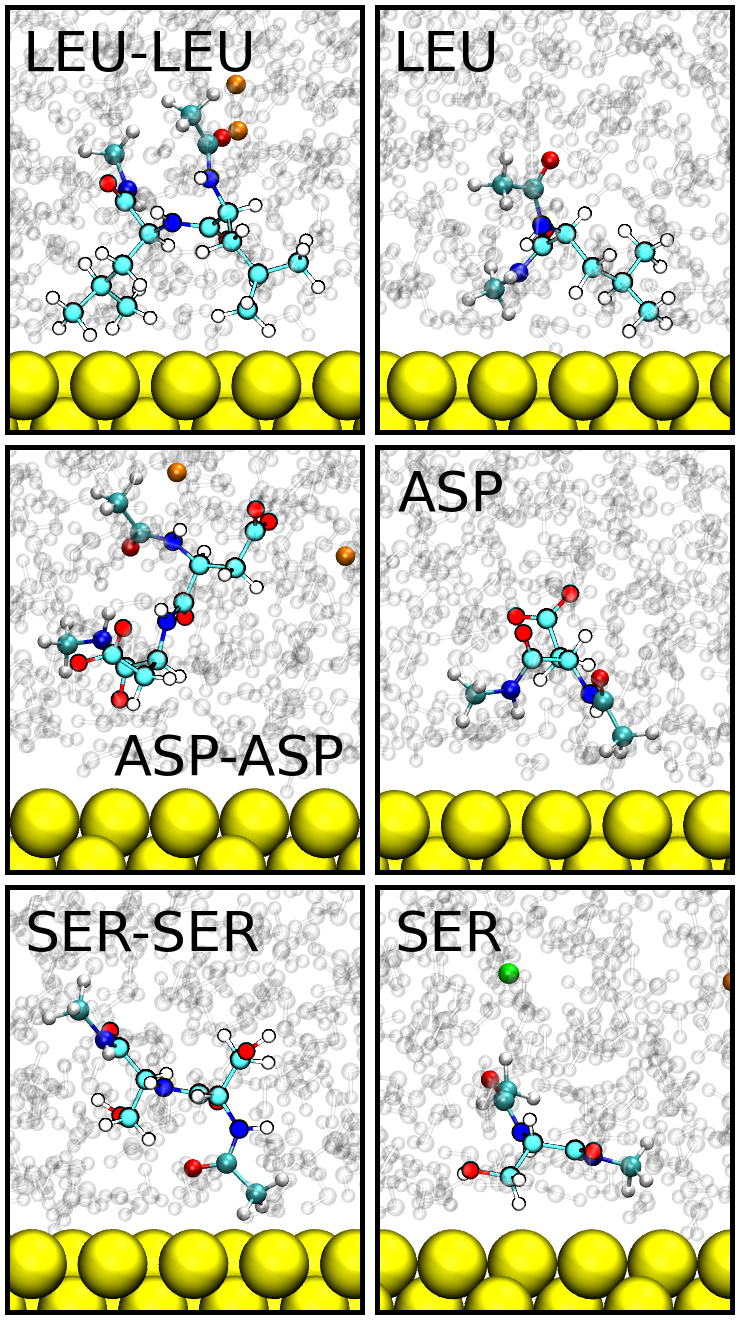}
\caption{The optimal conformations of selected homodipeptides and their
corresponding single amino acids (both with caps) studied with the FFB
force field. The spheres in cyan correspond to the carbon atoms, in red to
oxygen, in blue to nitrogen and in white to hydrogen. Amino acids are
shown in coats. The remaining atoms stand for caps. The separated spheres
indicate ions -- Cl$^{-}$ in lemon, Na$^{+}$ in orange. The dipeptides
are shown against the background of the water molecules.}
\label{bizzarri_dip_pic_a}
\end{center}
\end{figure}

\begin{figure}[ht]
\begin{center}
\includegraphics[scale=0.35]{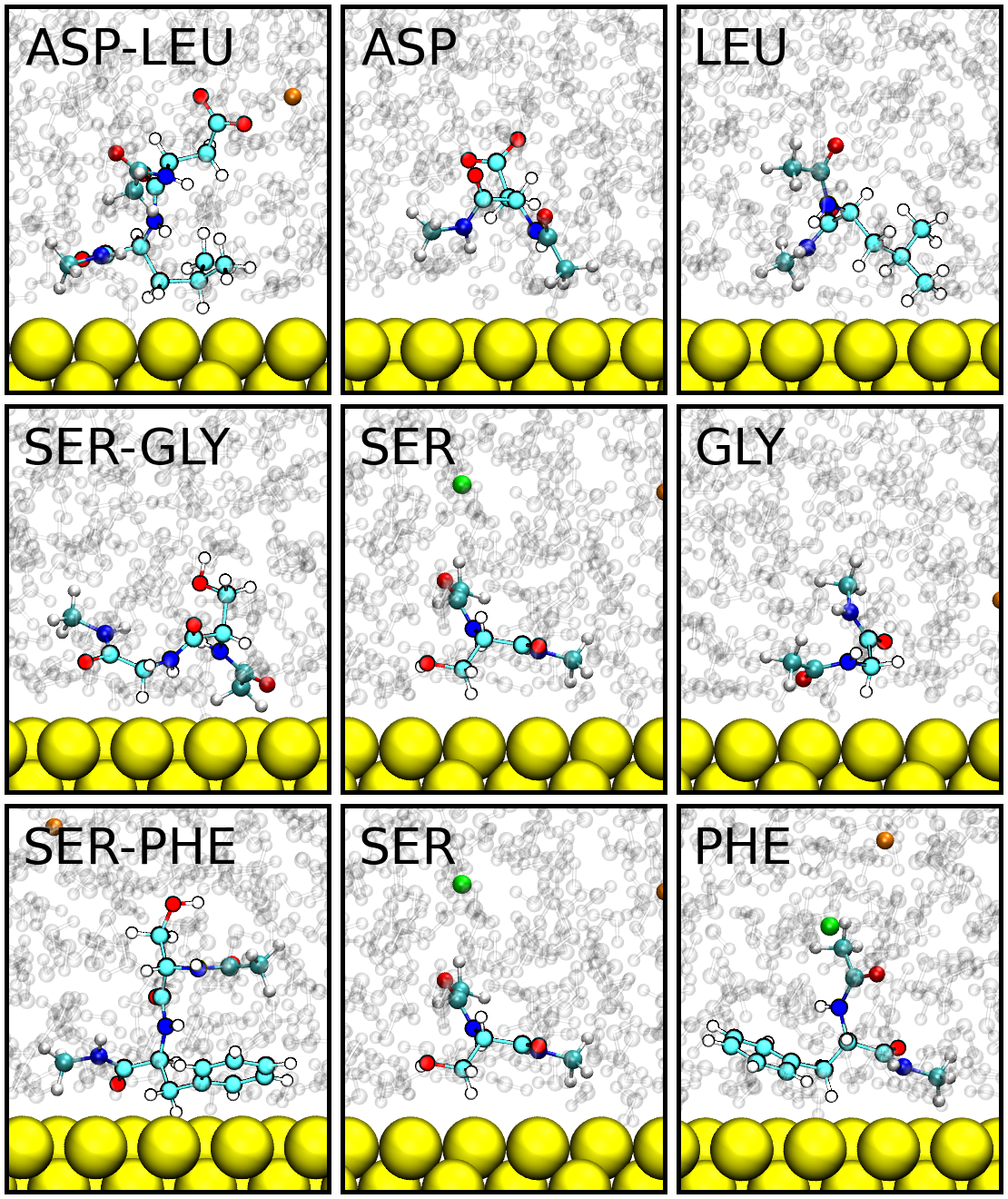}
\caption{Similar to Fig. \ref{bizzarri_dip_pic_a} but for the indicated heterodipeptides}
\label{bizzarri_dip_pic_b}
\end{center}
\end{figure}

Fig. \ref{bizzarri_dip_pmf} shows the PMF plots for 22 combinations
and compares them with the plots for the constituent single AAs. The binding
energies are summarized in Tab. \ref{bizzarri_dip_tab}.
The bond lengths, $\sigma$, for the dipeptides are generally larger 
than for the single AAs and the potential wells are wider. Both of these 
features simply reflect the bigger size of the molecules. 
Figs. \ref{bizzarri_dip_pic_a} and \ref{bizzarri_dip_pic_b}
show optimal conformations
for selected homo- and hetero-dipeptides respectively.
They demonstrate that the side chains
of the two AAs interact with one another which prevents adoption
of conformations that would be optimal individually.

\clearpage

More interestingly, we observe that when two strongly binding AAs combine 
(PHE--LEU, LEU--LEU, and PHE--PHE) then the resulting $\epsilon$ is 
bigger than each of the individual contributions but smaller than
the sum of them. 
For instance, LEU binds to gold through two hydrophobic
alkyl (methyl) groups (one in the cap) and LEU--LEU through three such groups. 
If one of them adopts a favorable conformation then the other
must point in the ways which are less favorable.
This can be brought out by considering the radius of
gyration, $R_g$, of the entities without the caps. For the single
LEU in the top right panel in Fig. \ref{bizzarri_dip_pic_a}, $R_g$ is 0.218 nm.
On the other hand, the first LEU in the dipeptide (
counting form the left in the left top panel)
has an $R_g$ of 0.208 nm and the second -- 0.238 nm.
The corresponding value of the radius of gyration around the $z$ axis,
$R_z$, are 0.191 nm, 0.167 nm, and 0.216 nm. Thus the 
the first LEU gets contracted and the second expanded on contact 
with the surface. As a result the binding energy for LEU--LEU is stronger than for LEU but 
is not twice as big. 
ASP binds with the surface weakly and the attraction is provided
primarily by one of the caps since the charged group is seen
to be repelled by the surface. For ASP-ASP the binding disappears
completely which indicates that the caps lead to an overestimation
of $\epsilon$ for single AAs.
A similar situation takes place for SER and SER--SER, except that
the dipeptide still binds. SER binds through the cap and the
nonpolar group of the side group (CH$_{2}$) whereas the polar 
group (OH) is attracted to water. The presence of the second SER
disallows adoption of such a conformation and binding is
implemented solely by the cap.

In heterodipeptides, the binding energy is often in between
the individual values of $\epsilon$. For instance, ASP--LEU
binds through the stronger coupling LEU, the presence of ASP
leads to a reduction in the strength of the coupling (the side 
group of ASP seeks water).
Similarly, GLY and SER--GLY both connect to gold through C$_{\alpha}$-H$_{2}$ 
and the cap but $\epsilon$ for the dipeptide is smaller than
for GLU. SER--PHE binds nearly as strongly as PHE because SER
is exposed to water without affecting the conformation of PHE
too much. 

\subsection{Adsorption of Tryptophan Cage to Gold}

We finally consider adsorption of tryptophan cage. This is a small
protein with the Protein Data Bank structure code of 1L2Y and it has no caps.
It comprises 20 AAs and none of them is a cysteine.
Its structure looks like a hairpin with one branch forming the
$\alpha$-helix (sites 2 through 8). There is a turn at site 10
and the 3/10-helix at 11 through 14. If one uses the hydropathy
scale of Kyte and Doolittle \cite{Kyte} then one infers that the
protein is overall hydrophilic (with the net hydropathy index of -19).

Instead of determining the PMF for 1L2Y, which is difficult to do
adequately, we simulate a free evolution of 20 ns and monitor events that 
result in adsorption. In the initial state, the CM
of the protein is placed at 2 nm above the gold surface  and in the
eight trajectories studied, for each force field, the initial
orientations are distinct. 
Unlike the situation with ZnO and ZnS \cite{nawrocki_2013,nawrocki_2014},
where the values of $\epsilon$ are weaker, adsorption is driven by the
AAs which we find to have the deepest single-AA PMFs and the adsorption
events are long lasting.

\begin{figure}[ht]
\begin{center}
\includegraphics[scale=0.4]{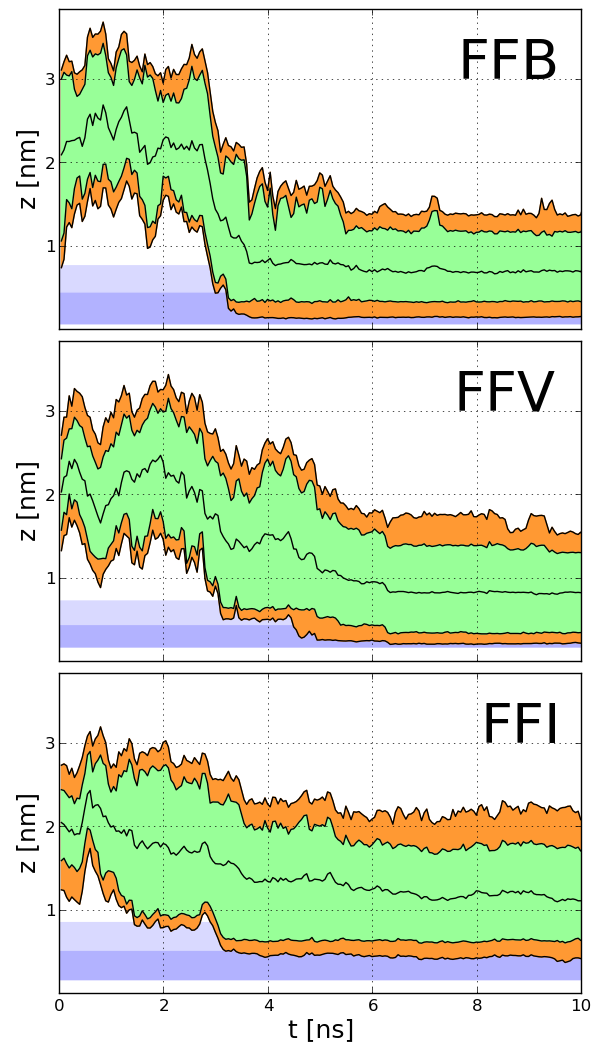}
\caption{Examples of the first 10 ns of evolution of 1L2Y in water above the three models of the surface of gold as indicated.        
The lines show, top to bottom, the instantaneous vertical positions of: 
the highest atom, the CM of the highest AA, the CM of the whole protein, 
the CM of the lowest AA, and the lowest atom. The identities of the highest
and lowest atoms vary in time. The horizontal layers at the bottom
indicate the first two layers in the water density profiles.
The first layer is in the darker shade of blue.
The continuation of the trajectories to 20 ns leaves the patterns essentially 
unchanged.}
\label{dis_pro_pan}
\end{center}
\end{figure}

Examples of the adsorption events are given in Fig. \ref{dis_pro_pan}
which shows vertical distances of characteristic locations in the
protein as a function of time. The central lines show the behavior
of the CM of the protein. Even though FFB provides couplings which
are the smallest in strength, the CM with this field comes closest to the
surface: it enters the second layer of water. For FFV, the CM comes
just above the second layer and for FFI it is substantially further away
which is consistent with the overall weaker couplings in FFI 
compared to FFV. Adsorption results in deformation and, judging by the
differences in vertical heights between the highest and lowest atoms,
the top panel corresponds to the biggest vertical flattening of the protein and
the bottom panel to the smallest, $i.e.$ to the largest vertical extension
(the initial extension depends on the orientation).
This is consistent with the values of $R_z$ of 0.573 nm, 0.545 nm, and 0.516 nm for
the top-to-bottom panels respectively.
On averaging over eight trajectories, 
we get $0.56\pm 0.02$ nm, $0.60\pm 0.02$ nm, and $0.56\pm 0.02$ nm
for FFB, FFV, and FFI respectively, with the standard radii of gyration
of $0.68 \pm 0.02$ nm, $0.60 \pm 0.02$, and  $0.63 \pm 0.02$ nm
whereas the native value (in the absence of gold) is about 0.78 nm.
The average $x$ and $y$ components are correspondingly:
0.74 and 0.55; 0.73 and 0.58; 0.73 and 0.60 nm.)
We conclude that, within the error bars, the three force fields
yield comparable distortions.

\clearpage

\begin{figure}[ht]
\begin{center}
\includegraphics[scale=0.4]{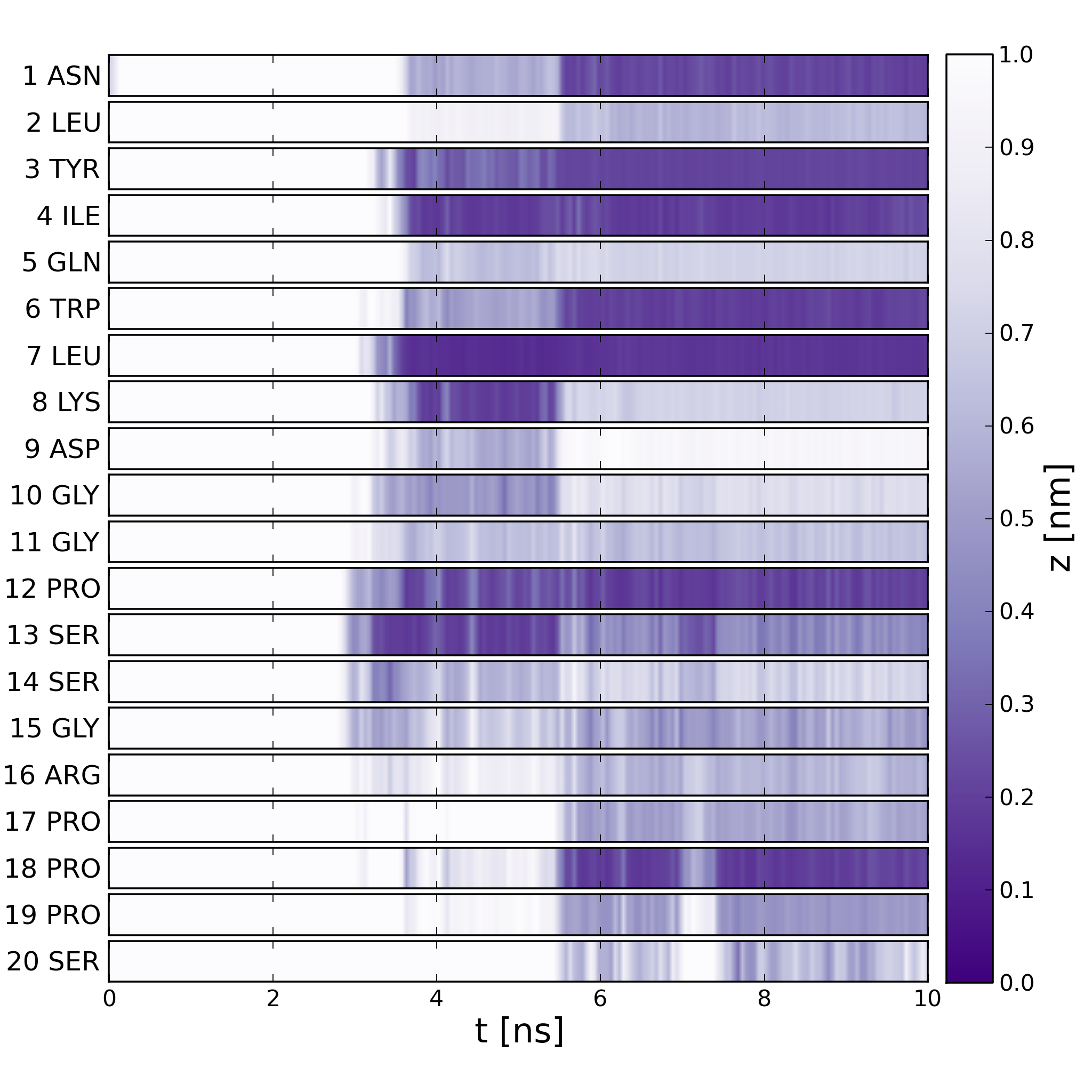}
\caption{The vertical positions of the lowest atoms of all AAs in the 1L2Y  in
the examples of free dynamics shown on Fig. \ref{dis_pro_pan} for the FFB model.}
\label{bizzarri_1l2y_ads}
\end{center}
\end{figure}

\begin{figure}[ht]
\begin{center}
\includegraphics[scale=0.3]{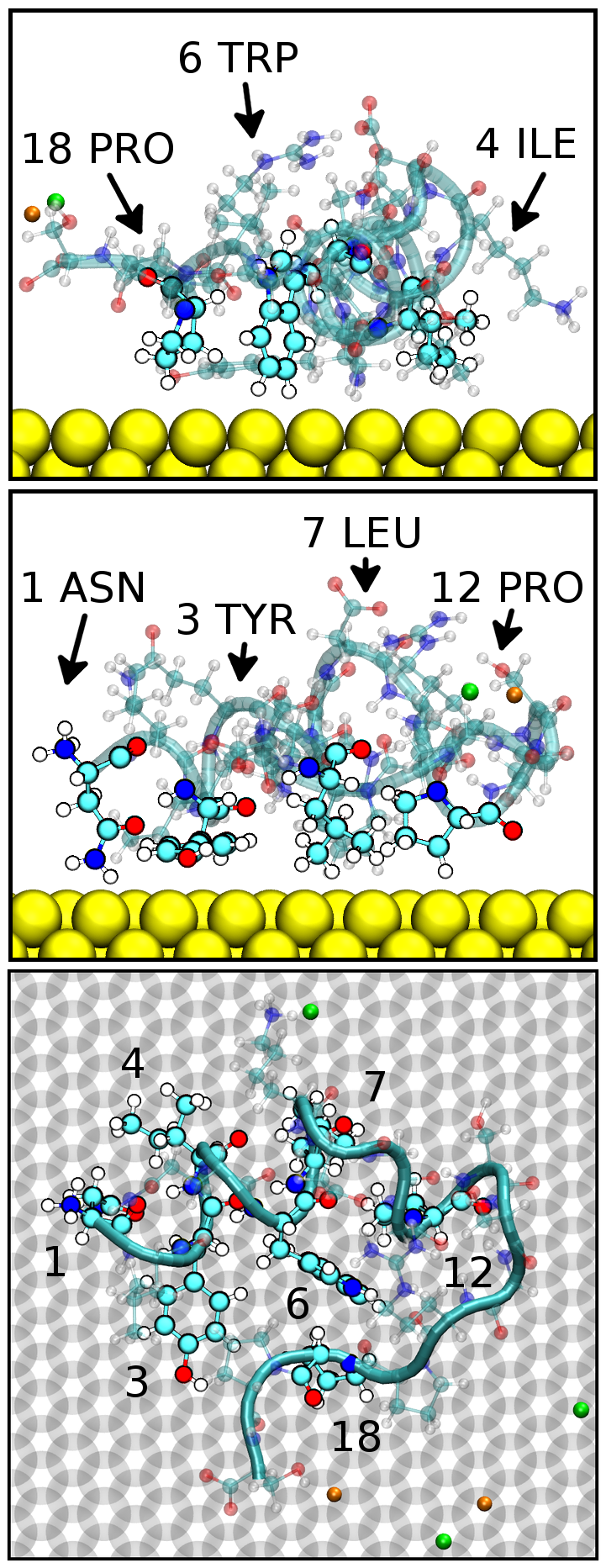}
\caption{The snapshot of adsorbed protein with the FFB model at  
8 ns in Fig. \ref{dis_pro_pan} and \ref{bizzarri_1l2y_ads}.
The conformation is shown from $x$, $y$ and $z$ directions, from top to bottom, respectively. 
The tube represents the full backbone in each panel.
The atoms of 7 AAs nearest the surface are shown in coats and indicated by the arrows. 
In the top view atoms of gold are transparent.
Molecules of water are not shown for clarity.}
\label{bizzarri_pro_pic}
\end{center}
\end{figure}

We now discuss the identity of the adsorbed AAs.
Fig. \ref{bizzarri_1l2y_ads}, for FFB, shows the vertical positions for each 
site in the sequence (the darker the tint, the deeper the lowest atom.
It is seen that the AAs that participate in the adsorption are
1 ASN, 3 TYR, 4 ILE, 6 TRP, 7 LEU, 12 PRO, and 18 PRO. 
All of them but 1 ASN are the strongest binders
listed in Tab. \ref{bizzarri_verde_aa_tab} for FFB, considering
that the protein has no GLU, CYS, PHE, and MET in its sequence.
Five of these binding AAs belong either to the $\alpha$- or 3/10-helix,
which correspond to regions of a lower configurational entropy
confirming the relevance of this notion for adsorption \cite{corni_2013}.
Figs. \ref{bizzarri_pro_pic} show the
adsorbed conformation from the top and two side-view panels.
It is seen that binding takes place through 1 ASN and six hydrophobic AAs.
Most of them, but ASN and TRP, adopt conformations that are similar
to the optimal ones obtained in the single-AA case.
With the exception of 18 PRO, the binding AAs belong to the helical part 
of the protein.

\begin{figure}[ht]
\begin{center}
\includegraphics[scale=0.32]{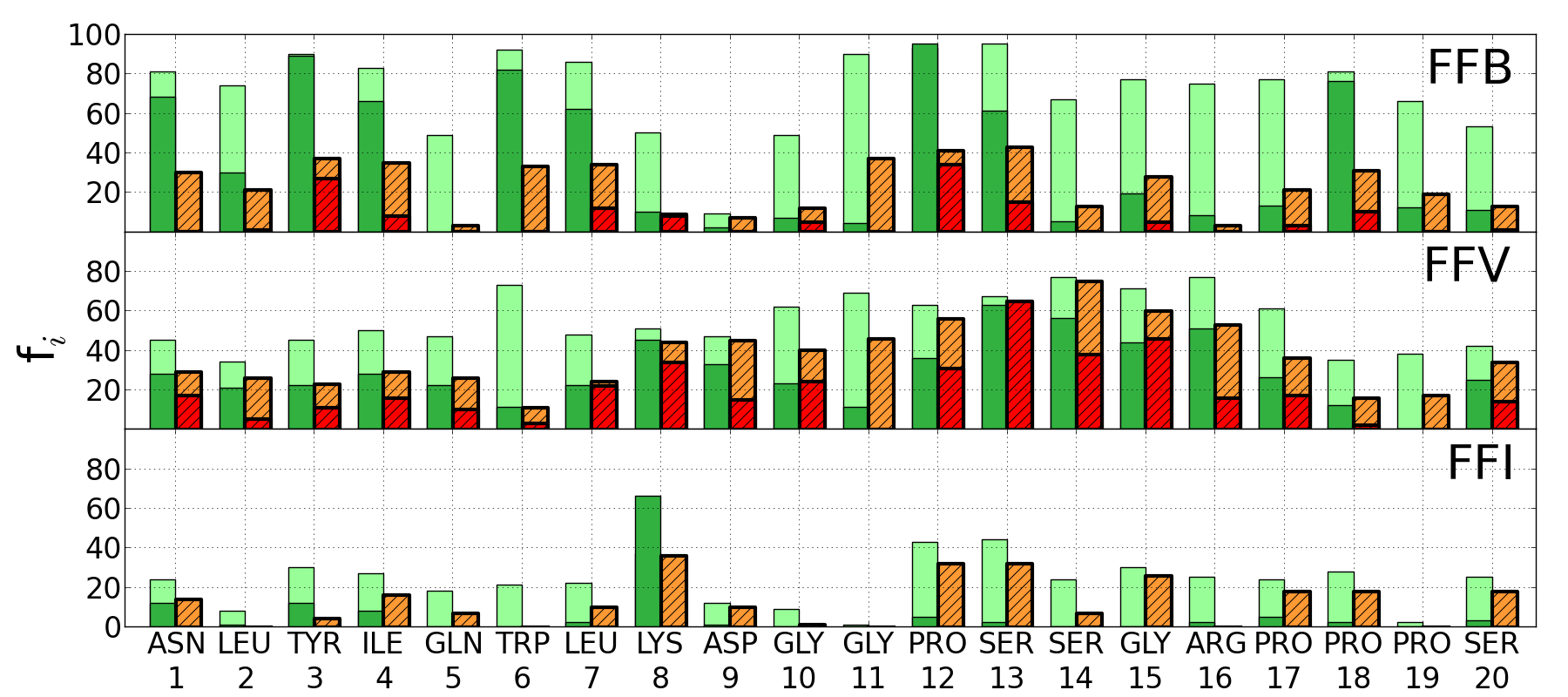}
\caption{Time of individual AAs bonding at the surface of Au 
expressed as a percentage of the protein total adsorption time
in eight trajectories for each force field.
The three models of the surfaces are considered, as indicated.
The darker colors (green and red) correspond to
the AAs entering the first water layer, whereas the lighter colors
(lime green and orange) to the AAs being either in  the first or second layers.
The uniformly shaded bins (on the left side) are used when the positions of the AAs 
are defined through their lowest atoms and the dashed bins (on the right side)
when the positions are defined through CM of the AAs.}
\label{pro_ads_sta}
\end{center}
\end{figure}

The statistics of the binding events  for the three force
fields are shown in Fig.  \ref{pro_ads_sta}.  The heights of the columns
indicate the percentage, $f_i$, of the total adsorption time of the protein
during which the $i$'th AA is either in layer I (darker shade) 
or below the upper boundary of layer II (lighter shade).
There are two kinds of columns for each AA: on the left the location of an
AA is defined through its lowest atom and on the right -- through its CM.
The adsorption starts to count when the lowest atom or the CM 
of the lowest AA crosses the
upper boundary of layer II. 

In the case of FFB, there are eight AAs which stay adsorbed for about 80\% of
the whole adsorption event (when using the description based on the location
of the lowest atom). These are the seven identified in 
Fig. \ref{bizzarri_pro_pic} and 13 SER which gets adsorbed in other trajectories. 
Except for 1 ASN, these AAs have the largest (available) values of $\epsilon$
in Table \ref{bizzarri_verde_aa_tab}. Note that the parameters in the table
refer to the CM of the AAs and not to the lowest atoms, but the fractions $f_i$
defined in terms of the CMs lead to similar conclusions.
The low value of $f_i$ for 9 ASP is
consistent with the low $\epsilon$ for ASP and, especially, the large $\sigma$.
ASN has $\epsilon$ that is comparable to that of ASP and a small $f_1$
would be expected to be found. However, the enhanced value of $f_1$ is due
to structural reasons: 1 ASN belongs to the helix and five AAs of the helix
bind to the surface strongly and force 1 ASN to follow suit.

The set of the well adsorbing AAs in the case of FFV is quite distinct
from that of FFB  and the $f_i$ are generally smaller
despite the generally stronger PMF-derived couplings.
The adsorption is mostly due to the two serins (13 and 14)
then 15 GLY, 16 ARG, 8 LYS and 6 TRP. We observe no obvious correlations
with the ranking of $\epsilon$ of Table \ref{bizzarri_verde_aa_tab} 

For FFI, the fractions $f_i$ become still smaller. Only 8 LYS gets $f_i$
exceeding 60\%. Most of the other AAs have $f_i$ of 20\% or less. This 
behavior appears to reflect the low level of specificity within FFI
and thus an effective interchangeability
-- from a trajectory to a trajectory -- of the binding AAs. There are
no correlations with the $\epsilon$'s in the table. For instance, 
TYR and TRP should bind the best but they do not. Furthermore, 17 PRO
has comparable $f_i$ to 18 PRO, but 19 PRO does not bind. Similarly,
$f_i$ for 15 GLY  is about three times as big as $f_i$ for 10 GLY
with $f_i$ for 11 GLY being negligible. This may reflect an insufficient
statistics to some extent, but is more likely due to the role
of the whole structure in determining which segments bind and which do not.
With high specificity, like in FFB, the role of the structure gets
diminished.

Simulations by Hoefling {\it et al.} \cite{hoefling_2010_b} for
single capped AAs described within FFI, indicate that the AAs can
come quite close to the surface during a free evolution.
We have confirmed this finding for ASP, MET, PHE, and TYR
(for which the PMFs in ref. \cite{hoefling_2010_b} have been shown
explicitely).
However, when these AAs are parts of the protein then they do not
penetrate the first water layer at all, as illustrated in the bottom
panel of Fig. \ref{dis_pro_pan}. It is statistically improbable
for several connected AAs to go through layer I which can also be
stated as an existence of correlations between the AAs.
The contributing circumstances are that conformations of AAs in the
protein are not the same as when in isolation and that the calculations
of the PMFs involve making averages over the lateral positions and thus
do not refer to a particular $x-y$ coordinates.

\clearpage

\section{Conclusions}

Generally, it is not surprising that results of molecular dynamics
simulations depend on the details of the force field. 
Such dependence has been documented for proteins even
without any interactions with solids \cite{Matthes,Piana,Karttunen}.
However, in the
case of aqueous biomolecules interacting with gold the differences between
the three force fields considered are profound and qualitative in nature.
The differences are related to the behavior of the molecules of water
just near the surface  that show in the density profiles and orientation
of the molecules. In terms of the average coupling strengths,
the hydrophobic FFB is the weakest and the non-polarizable hydrophilic
FFV is the strongest. We have checked that the ranking of the FFV-based 
couplings mostly agrees with that obtained for the analogues of the
side chains (or their fragments). The FFB couplings,
with the exception of the covalently binding CYS, are, on average,
an order stronger than those obtained for the semiconducting ZnO.
However, the strongest couplings are larger only by a factor 
between 1.84 and 4.78, depending on the choice of the 
exposed surface of ZnO. 

In terms of the specificities, FFB leads to the
largest sensitivity (49\%) to the identity of amino acids and the
polarizable FFI to the smallest (25\%). On the other hand, specificities
for ZnO are much larger -- they range between 68\% and 154\%, depending  
on the surface. Despite the weaker FFB-based AA-Au couplings than
for the hydrophilic force fields, the hydrophobic surface actually
binds proteins better statistically (if no CYS is involved) 
because of the stronger specificity and a possibility of a closer approach
to the surface through water. At the same time, the levels of distortion
of the protein are similar for the three force fields.

If one asks whether one can predict which protein or peptide may bind
to gold based on the single AA binding strengths, then the answer is
that this may work fairly well in the hydrophobic case even though
our results for the dipeptides do not support a simple additivity
of the single-AA contributions. In the hydrophilic cases, it seems
better to gather statistics of the frequency with which, say, the 
lowest atoms come to the first two layers of water and derive
effective couplings via the Boltzmann factors. An alternative is to
use the PMF-based single AA-potentials but with an extra softly
repulsive potential that prevents a too close approach to the surface.
The couplings determined in this paper can be employed in coarse simulations,
for instance of the kind proposed in refs. \cite{pandey_2009,feng_2011}.

\vspace{0.5cm}

{\Large \textbf{Acknowledgement}}

\vspace{0.5cm}

Discussions with D. Elbaum, J. Grzyb and B. R\'o\.zycki are warmly appreciated.
This work has been supported by
the Polish National Science Centre Grants No. 2011/01/B/ST3/02190 
as well as by the European Union within European Regional Development Fund,
through Innovative Economy grant (POIG.01.01.02-00-008/08).
The local computer resources were financed by the European Regional Development Fund
under the Operational Programme Innovative Economy NanoFun POIG.02.02.00-00-025/09.
We appreciate help of A. Koli{\'n}ski and A. Liwo in providing additional
computer resources such as at the Academic Computer Center in Gda\'nsk.


\newpage

\begin{centering}
{{\large {\bf Supplementary Information} }\\
\vspace{0.5cm}
\LARGE \bf Aqueous Amino Acids and Proteins Near the Surface of Gold 
       in Hydrophilic and Hydrophobic Force Fields}\\

\vspace{0.5cm}

{\large Grzegorz Nawrocki and Marek Cieplak}\\
{\small Institute of Physics, Polish Academy of Sciences, Al. Lotnik\'ow 32/46, 02-668 Warsaw, Poland}\\
\end{centering}

\vspace{1cm}

\begin{figure}[ht]
\begin{center}
\includegraphics[scale=0.25]{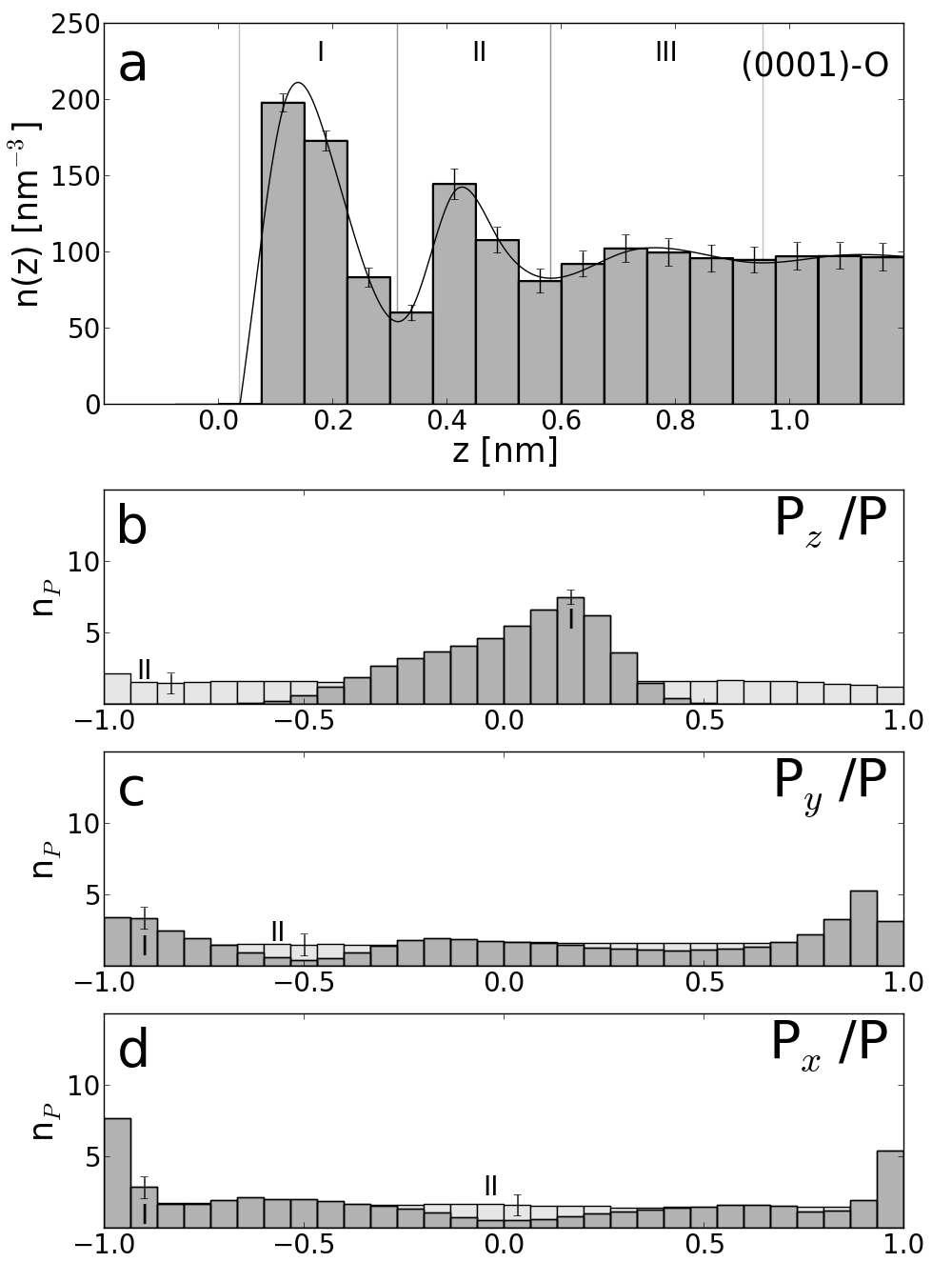}
\caption{The distributions of density and  water polarization  
above the (0001)-O) surface of ZnO.} 
\label{0001O}
\end{center}
\end{figure}

\begin{figure}[ht]
\begin{center}
\includegraphics[scale=0.25]{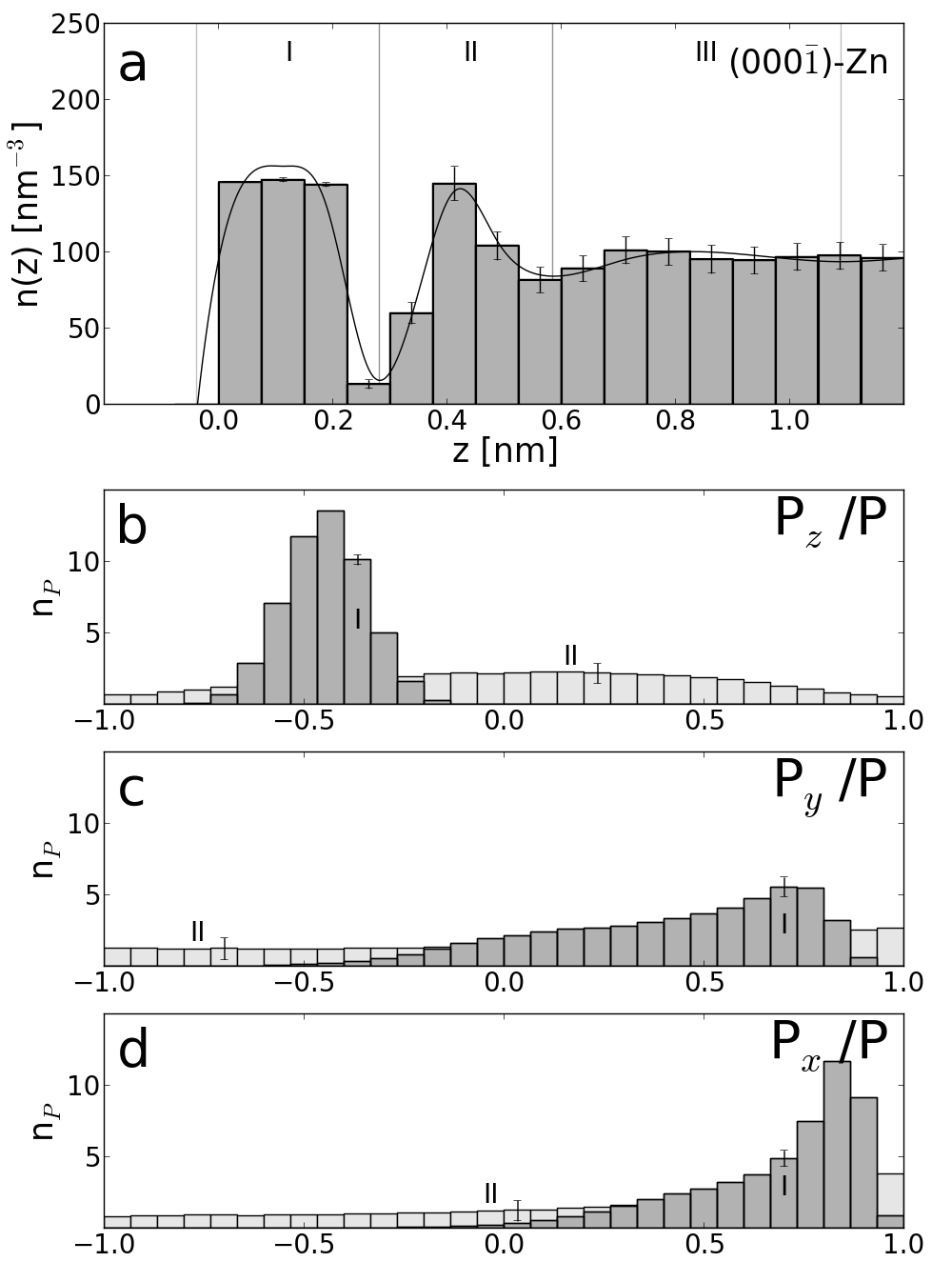}
\caption{Similar to Fig. \ref{0001O} but above the (000$\bar{1}$)-Zn surface of ZnO.}
\label{0001Zn}
\end{center}
\end{figure}

\begin{figure}[ht]
\begin{center}
\includegraphics[scale=0.25]{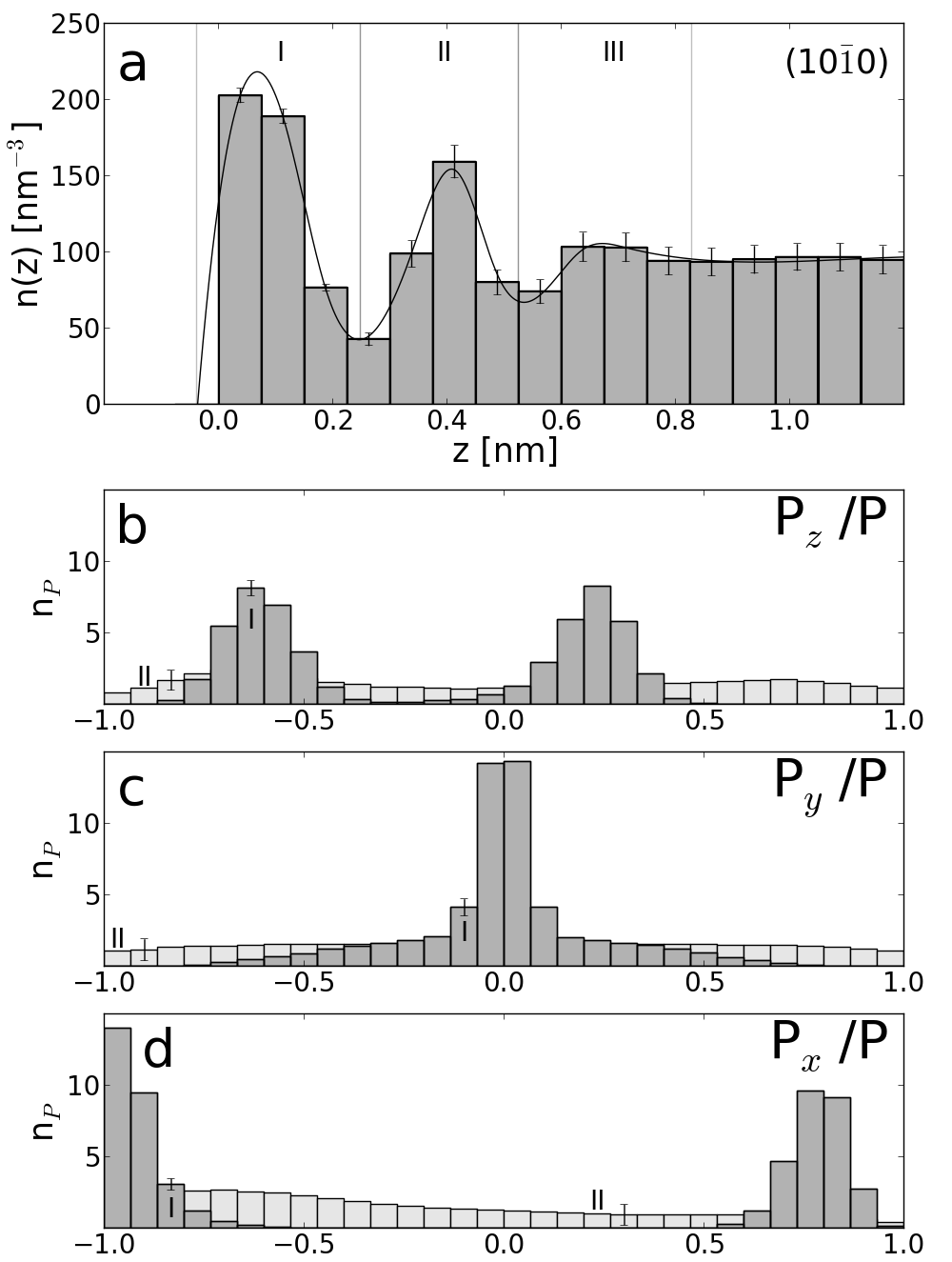}
\caption{Similar to Fig. \ref{0001O} but above the (10$\bar{1}$0) surface of ZnO.}
\label{1010}
\end{center}
\end{figure}

\begin{figure}[ht]
\begin{center}
\includegraphics[scale=0.25]{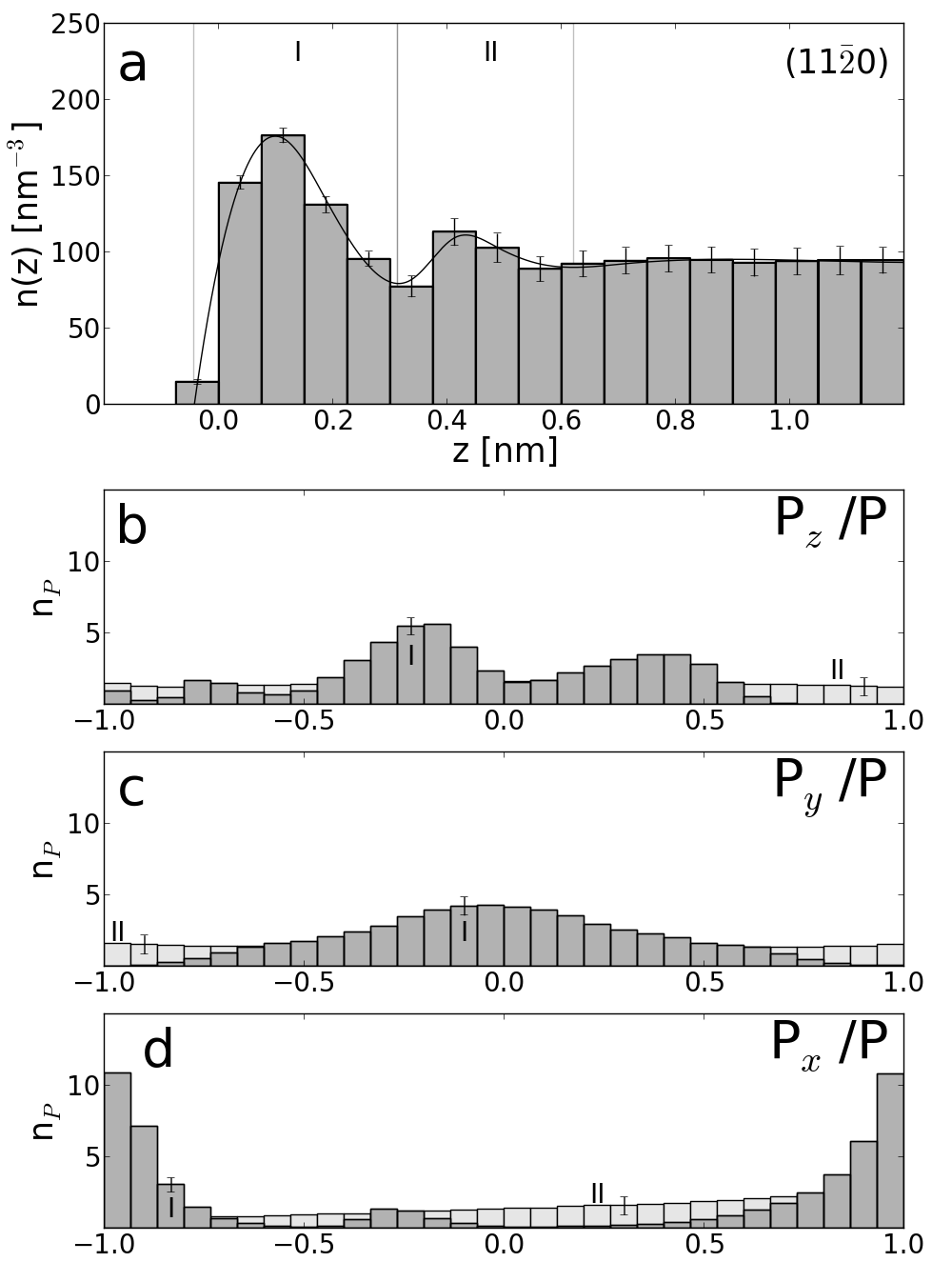}
\caption{Similar to Fig. \ref{0001O} but above the (11$\bar{2}$0) surface of ZnO.}
\label{1120}
\end{center}
\end{figure}

In order to discuss the properties of water near gold, it is useful
to provide a background of other possible behaviors, such as near ZnO.
This system has been studied in
ref. \cite{nawrocki_2013} but the distributions of the polarization
vector have not been shown there.
We have considered four cuts of the bulk crystal, two polar ((0001)-O, (000$\bar{1}$)-Zn)
and two nonpolar ((10$\bar{1}$0), (11$\bar{2}$0). The corresponding density and
polarization profiles are shown in Figs. \ref{0001O}, \ref{0001Zn},
\ref{1010}, and \ref{1120} respectively.

\begin{figure}[ht]
\begin{center}
\includegraphics[scale=0.4]{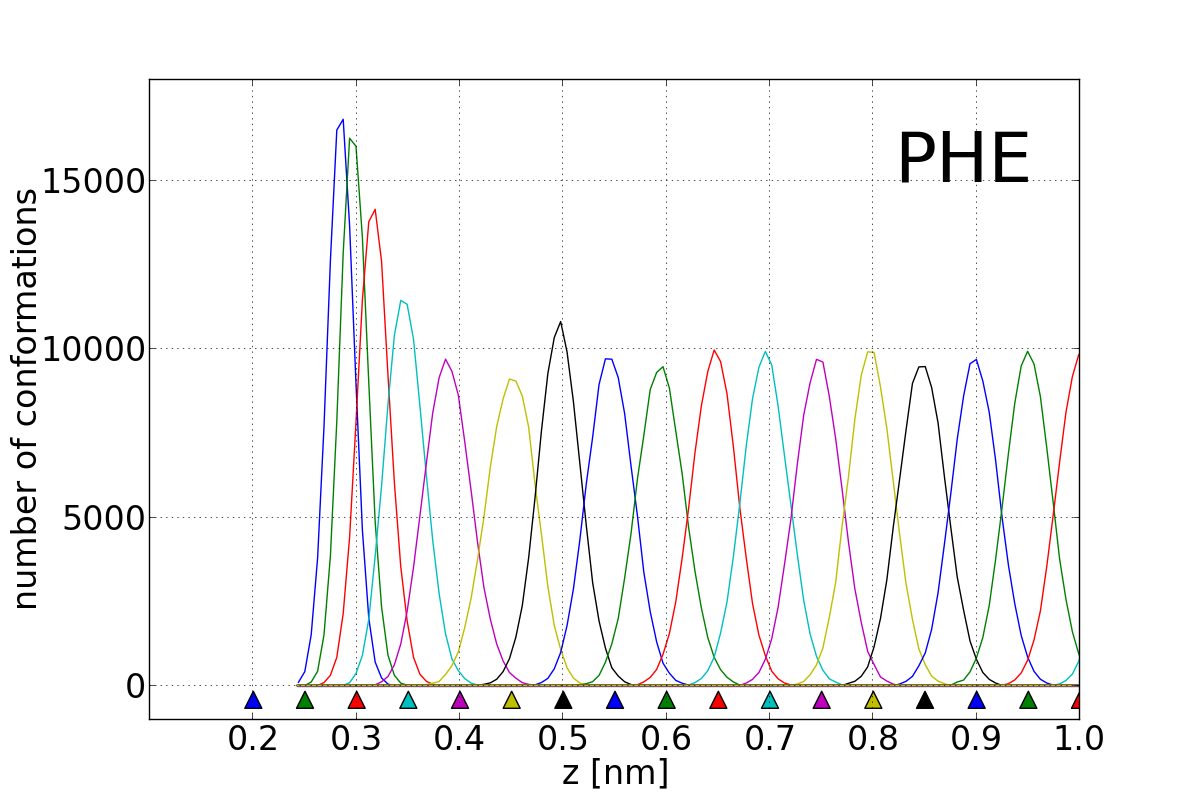}
\caption{Histogram of the number of conformations of phenylalanine above
the gold surface in FFB
as obtained through the umbrella sampling method.
The different colors correspond to various simulation windows.
In each simulation, the AA is restrained by the umbrella biasing potential
to different selected values of $z$
that are marked as triangles in a color corresponding to the simulation
window.
For instance, the maximum of the distribution for z=0.70 nm is close to
the set value, indicating a weak impact of the surface.
For 0.35 and 0.40 nm, the maxima are shifted toward the surface due to the
attraction.
However, for 0.20, 0.25 and 0.30 nm, the maxima are shifted away from the
surface due to the overlaping.
Generally, a large shift away from the set value comes with a narrower and
taller distribution.
A wide distribution, as for z=0.70 nm, suggests that the AA is attracted
by the surface weakly.}
\label{histo}
\end{center}
\end{figure}

Fig. \ref{histo} illustrates the umbrella sampling method. 
It shows the distributions of the
numbers of conformations at various sampling locations for PHE
considered within the FFB approach.

\begin{figure}[h!]
\centering
\subfigure[]{
\includegraphics[scale=0.4]{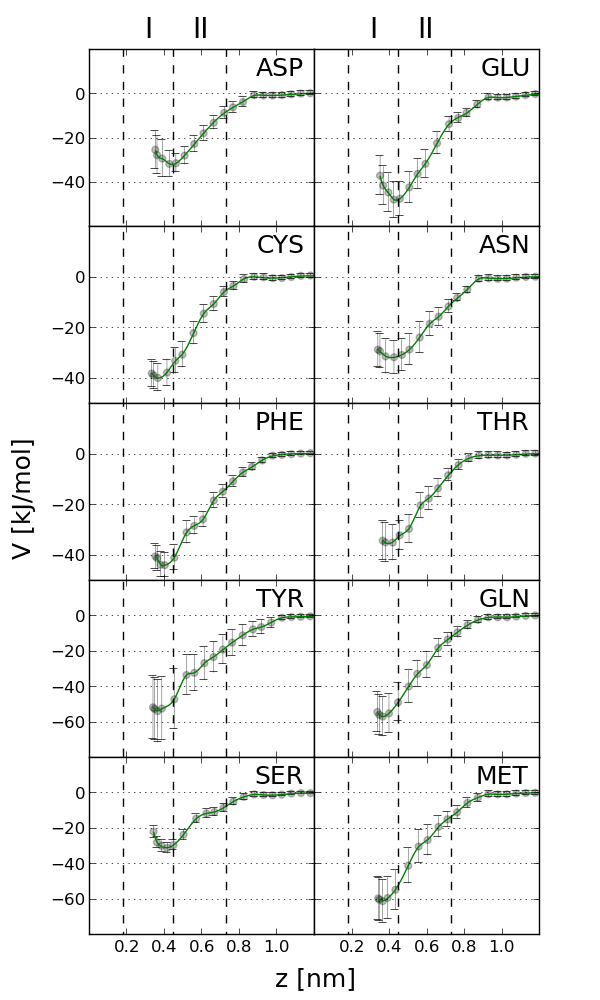}}
\subfigure[]{
\includegraphics[scale=0.4]{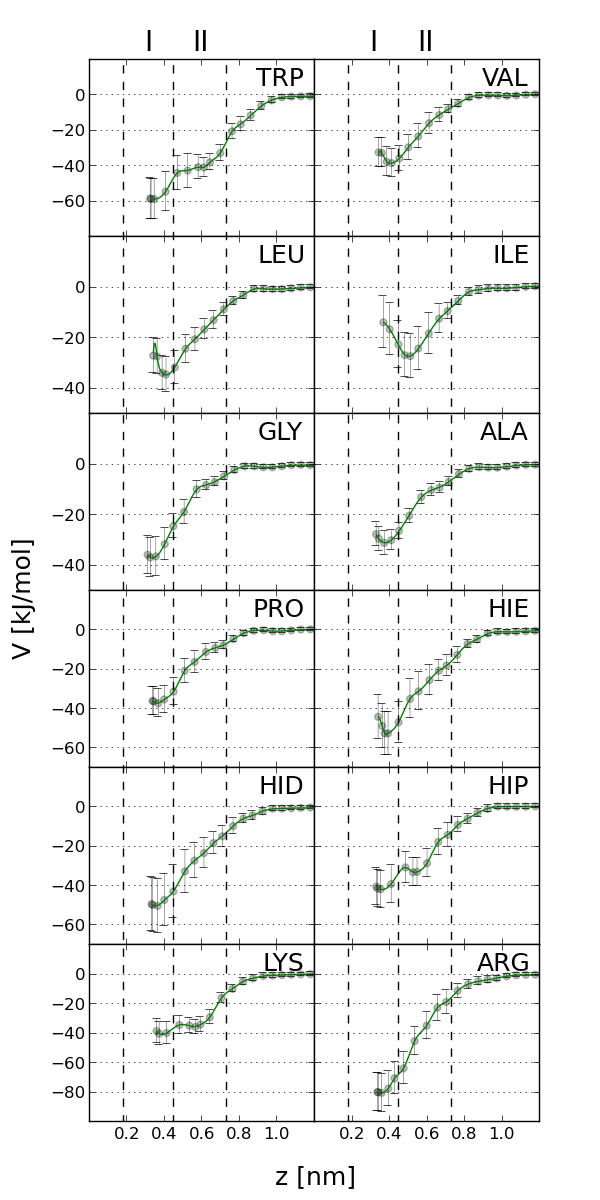}}
\caption{The PMF for single AAs with caps in water solutions as obtained by
using the hydrophilic FFV model. The depth of the lowest negative minimum 
in $V(z)$ defines $\epsilon$ and its location -- $\sigma$.}
\label{verde_aa_pmf}
\end{figure}

\begin{figure}[h!]
\centering
\subfigure[]{
\includegraphics[scale=0.4]{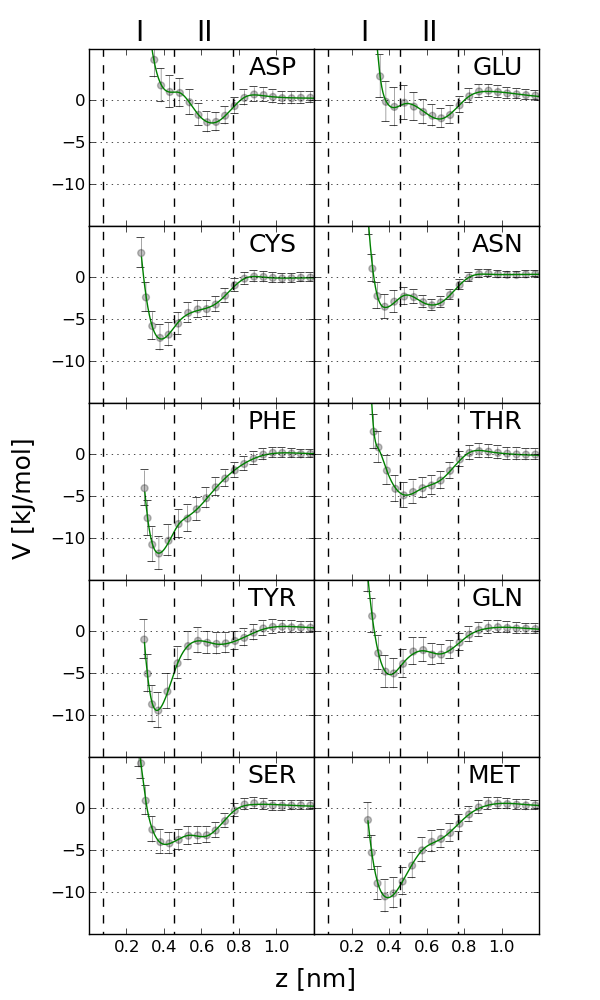}}
\subfigure[]{
\includegraphics[scale=0.4]{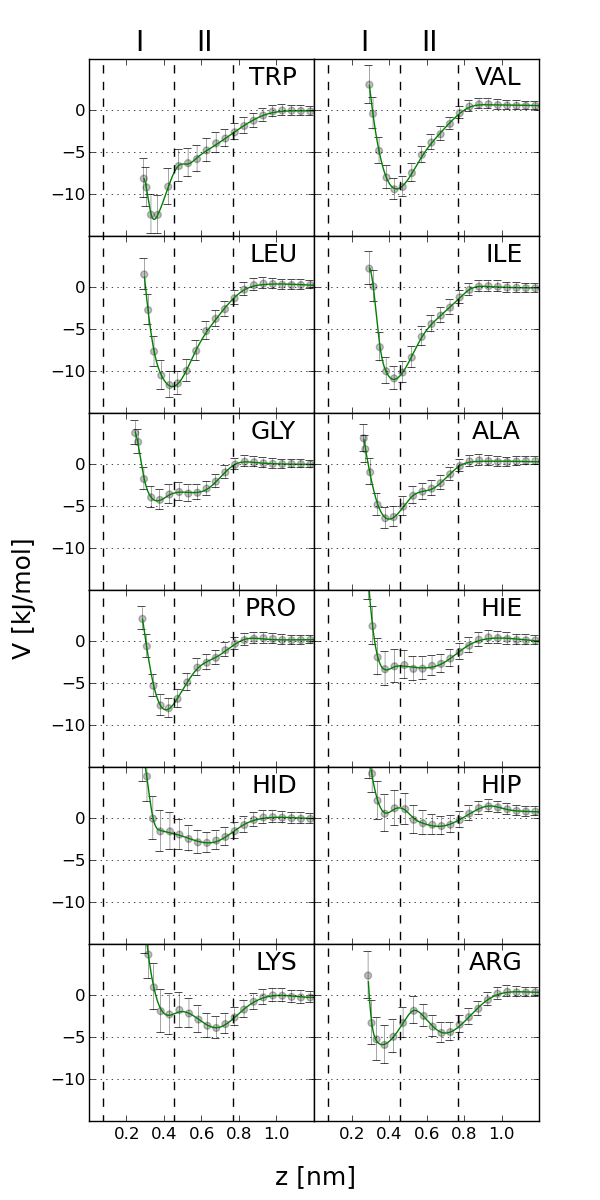}}
\caption{Similar to Fig. \ref{verde_aa_pmf} but for the hydrophobic FFB model.}
\label{bizzarri_aa_pmf}
\end{figure}

Figs. \ref{verde_aa_pmf} and \ref{bizzarri_aa_pmf} show the PMFs for
the twenty amino acids as determined within the FFV and FFB force fields
respectively.

\begin{figure}[h!]
\centering
\includegraphics[scale=0.42]{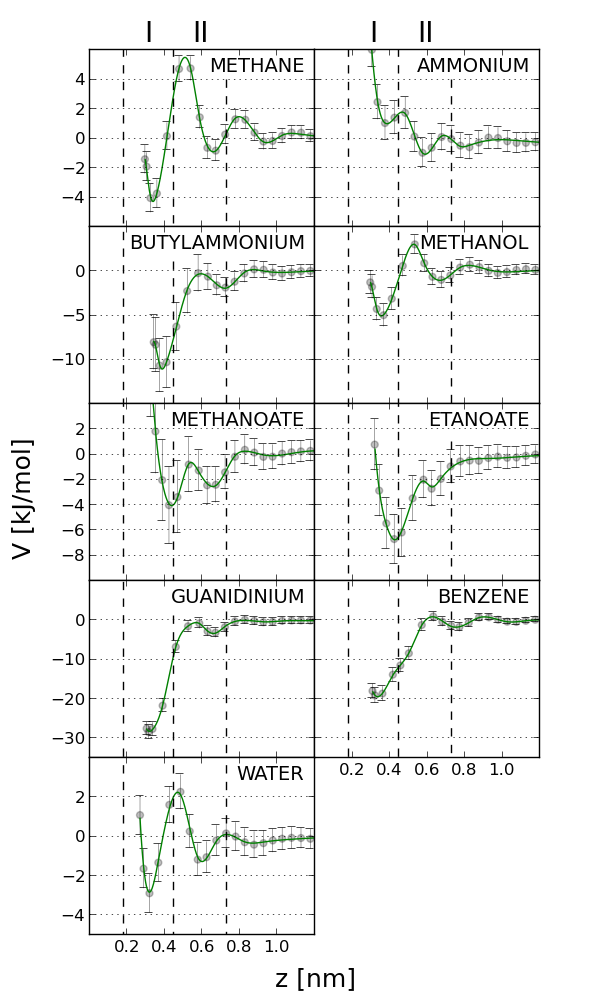}
\caption{The PMF for the analogues of the AA side chains and water molecule 
in water solutions in the hydrophilic FFV model.}
\label{verde_an_pmf}
\end{figure}

Fig. \ref{verde_an_pmf} shows the PMF for the analogues of the AAs side 
chains discussed
in the main text and for the hydrophilic FFV model.

\end{document}